\documentclass[twocolumn]{aastex631}

\usepackage{hyperref}

\shorttitle{Asphericity in SN~2023ixf}
\shortauthors{Singh et al.}

\begin{document}

\title{Unravelling the asphericities in the explosion and multi-faceted circumstellar matter of SN 2023ixf}

\author[0000-0003-2091-622X]{Avinash Singh}
\affiliation{Hiroshima Astrophysical Science Centre, Hiroshima University, 1-3-1 Kagamiyama, Higashi-Hiroshima, Hiroshima 739-8526, Japan}
\affiliation{Department of Astronomy, The Oskar Klein Center, Stockholm University, AlbaNova University Center, SE 106 91 Stockholm, Sweden}

\author[0000-0002-0525-0872]{Rishabh Singh Teja}
\affiliation{Indian Institute of Astrophysics, II Block, Koramangala, Bengaluru-560034, Karnataka, India}
\affiliation{Pondicherry University, R.V. Nagar, Kalapet, Pondicherry-605014, UT of Puducherry, India}

\author[0000-0003-1169-1954]{Takashi J. Moriya}
\affiliation{National Astronomical Observatory of Japan, National Institutes of Natural Sciences, 2-21-1 Osawa, Mitaka, Tokyo 181-8588, Japan}
\affiliation{Graduate Institute for Advanced Studies, SOKENDAI, 2-21-1 Osawa, Mitaka, Tokyo 181-8588, Japan}
\affiliation{School of Physics and Astronomy, Monash University, Clayton, Victoria 3800, Australia}

\author[0000-0003-2611-7269]{Keiichi Maeda}
\affiliation{Department of Astronomy, Kyoto University, Kitashirakawa-Oiwake-cho, Sakyo-ku, Kyoto 606-8502, Japan}

\author[0000-0001-6099-9539]{Koji S Kawabata}
\affiliation{Hiroshima Astrophysical Science Centre, Hiroshima University, 1-3-1 Kagamiyama, Higashi-Hiroshima, Hiroshima 739-8526, Japan}

\affiliation{Department of Physics, Graduate School of Advanced Science and Engineering, Hiroshima University, 1-3-1 Kagamiyama, Higashi-Hiroshima, Hiroshima 739-8526, Japan}

\author[0000-0001-8253-6850]{Masaomi Tanaka}
\affiliation{Astronomical Institute, Tohoku University, Aoba, Sendai 980-8578, Japan}

\author{Ryo Imazawa}
\affiliation{Department of Physics, Graduate School of Advanced Science and Engineering, Hiroshima Astrophysical Science Centre, Hiroshima University, 1-3-1 Kagamiyama, Higashi-Hiroshima, Hiroshima 739-8526, Japan}

\author{Tatsuya Nakaoka}
\affiliation{Hiroshima Astrophysical Science Centre, Hiroshima University, 1-3-1 Kagamiyama, Higashi-Hiroshima, Hiroshima 739-8526, Japan}

\author[0000-0002-3884-5637]{Anjasha Gangopadhyay}
\affiliation{Hiroshima Astrophysical Science Centre, Hiroshima University, 1-3-1 Kagamiyama, Higashi-Hiroshima, Hiroshima 739-8526, Japan}

\author[0000-0001-9456-3709]{Masayuki Yamanaka}
\affiliation{Amanogawa Galaxy Astronomy Research Center (AGARC), Graduate School of Science and Engineering, Kagoshima University, 1-21-35 Korimoto, Kagoshima, Kagoshima 890-0065, Japan}

\author[0000-0002-7942-8477]{Vishwajeet Swain}
\affiliation{Department of Physics, Indian Institute of Technology Bombay, Powai, Mumbai 400076}

\author[0000-0002-6688-0800]{D.K. Sahu}
\affiliation{Indian Institute of Astrophysics, II Block, Koramangala, Bengaluru-560034, Karnataka, India}

\author[0000-0003-3533-7183]{G.C. Anupama}
\affiliation{Indian Institute of Astrophysics, II Block, Koramangala, Bengaluru-560034, Karnataka, India}

\author[0000-0001-7225-2475]{Brajesh Kumar}
\affiliation{South-Western Institute for Astronomy Research, Yunnan University, Kunming, Yunnan 650504, People's Republic of China}

\author[0000-0002-4989-6253]{Ramya M. Anche}
\affiliation{Steward Observatory, University of Arizona, 933N Cherry Avenue, Tucson, AZ 85721, USA}

\author{Yasuo Sano}
\affiliation{Observation and Data Center for Cosmosciences, Faculty of Science, Hokkaido University,
Kita-ku, Sapporo-shi, Hokkaido 060-0810, Japan}
\affiliation{Nayoro Observatory, 157-1 Nisshin, Nayoro-shi, Hokkaido 096-0066, Japan}

\author{A. Raj}
\affiliation{Indian Centre for Space Physics, 466 Barakhola, Netai Nagar, Kolkata, 700099, West Bengal, India}

\author{V. K. Agnihotri}
\affiliation{Cepheid Observatory, India}

\author[0000-0002-6112-7609]{Varun Bhalerao}
\affiliation{Department of Physics, Indian Institute of Technology Bombay, Powai, Mumbai 400076}

\author{D. Bisht}
\affiliation{Indian Centre for Space Physics, 466 Barakhola, Netai Nagar, Kolkata, 700099, West Bengal, India}

\author{M. S. Bisht}
\affiliation{Indian Centre for Space Physics, 466 Barakhola, Netai Nagar, Kolkata, 700099, West Bengal, India}

\author{K. Belwal}
\affiliation{Indian Centre for Space Physics, 466 Barakhola, Netai Nagar, Kolkata, 700099, West Bengal, India}

\author{S. K. Chakrabarti}
\affiliation{Indian Centre for Space Physics, 466 Barakhola, Netai Nagar, Kolkata, 700099, West Bengal, India}

\author{Mitsugu Fujii}
\affiliation{Fujii Kurosaki Observatory, 4500 Kurosaki, Tamashima, Kurashiki, Okayama 713-8126, Japan}

\author{Takahiro Nagayama}
\affiliation{Graduate School of Science and Engineering, Kagoshima University, 1-21-35 Korimoto, Kagoshima, Kagoshima 890-0065, Japan}

\author[0000-0002-5277-568X]{Katsura Matsumoto}
\affiliation{Astronomical Institute, Osaka Kyoiku University, Kashiwara-shi, Osaka 582-8582, Japan}

\author{Taisei Hamada}
\affiliation{Department of Physics, Graduate School of Advanced Science and Engineering, Hiroshima University, Kagamiyama, 1-3-1 Higashi-Hiroshima, Hiroshima 739-8526, Japan}
\affiliation{Hiroshima Astrophysical Science Centre, Hiroshima University, 1-3-1 Kagamiyama, Higashi-Hiroshima, Hiroshima 739-8526, Japan}

\author[0000-0002-4540-4928]{Miho Kawabata}
\affiliation{Nishi-Harima Astronomical Observatory, Center for Astronomy, University of Hyogo, 407-2 Nishigaichi, Sayo-cho, Sayo, Hyogo, 679-5313, Japan}

\author[0000-0002-4870-9436]{Amit Kumar}
\affiliation{Department of Physics, University of Warwick, Gibbet Hill Road, Coventry CV4 7AL, UK}

\author[0009-0008-6428-7668]{Ravi Kumar}
\affiliation{Department of Aerospace Engineering, Indian Institute of Technology Bombay, Powai, Mumbai 400076}

\author[0000-0002-8543-6406]{Brian K. Malkan}
\affiliation{Department of Astronomy, Case Western Reserve University, 10900 Euclid Avenue, Cleveland, OH 44106, USA}

\author[0000-0002-5083-3663]{Paul Smith}
\affiliation{Steward Observatory, University of Arizona, 933N Cherry Avenue, Tucson, AZ 85721, USA}

\author{Yuta Sakagami}
\affiliation{Faculty of Agriculture, Kyoto University, Kitashirakawa-Oiwake-cho, Sakyo-ku, Kyoto 606-8502, Japan}

\author[0000-0002-8482-8993]{Kenta Taguchi}
\affiliation{Department of Astronomy, Kyoto University, Kitashirakawa-Oiwake-cho, Sakyo-ku, Kyoto 606-8502, Japan}

\author[0000-0001-8537-3153]{Nozomu Tominaga}
\affiliation{National Astronomical Observatory of Japan, National Institutes of Natural Sciences, 2-21-1 Osawa, Mitaka, Tokyo 181-8588, Japan}
\affiliation{Astronomical Science Program, Graduate Institute for Advanced Studies, SOKENDAI, 2-21-1 Osawa, Mitaka, Tokyo 181-8588, Japan}
\affiliation{Department of Physics, Faculty of Science and Engineering, Konan University, 8-9-1 Okamoto, Kobe, Hyogo 658-8501, Japan}

\author{Arata Watanabe}
\affiliation{Faculty of Science, Kyoto University, Kitashirakawa-Oiwake-cho, Sakyo-ku, Kyoto 606-8502, Japan}

\correspondingauthor{Avinash Singh, Rishabh Singh Teja}
\email{avinash21292@gmail.com, rsteja001@gmail.com}

\begin{abstract}

We present a detailed investigation of photometric, spectroscopic, and polarimetric observations of the Type II SN~2023ixf. Earlier studies have provided compelling evidence for a delayed shock breakout from a confined dense circumstellar matter (CSM) enveloping the progenitor star. The temporal evolution of polarization in SN~2023ixf revealed three distinct peaks in polarization evolution at 1.4 d, 6.4 d, and 79.2 d, indicating an asymmetric dense CSM, an aspherical shock front and clumpiness in the low-density extended CSM, and an aspherical inner ejecta/He-core. SN 2023ixf displayed two dominant axes, one along the CSM-outer ejecta and the other along the inner ejecta/He-core, showcasing the independent origin of asymmetry in the early and late evolution. The argument for an aspherical shock front is further strengthened by the presence of a high-velocity broad absorption feature in the blue wing of the Balmer features in addition to the P-Cygni absorption post 16\,d. Hydrodynamical light curve modeling indicated a progenitor mass of 10 $\rm M_{\odot}$ with a radius of 470 $\rm R_{\odot}$ and explosion energy of $2\times 10^{51}\,\mathrm{erg}$, along with 0.06 $\rm M_{\odot}$ of $^{56}\,\mathrm{Ni}$, though these properties are not unique due to modeling degeneracies. The modeling also indicated a two-zone CSM: a confined dense CSM extending up to $5 \times 10^{14}\,\mathrm{cm}$, with a mass-loss rate of 10$^{-2}$ $\rm M_{\odot}\ yr^{-1}$ and an extended CSM spanning from $5 \times 10^{14}\,\mathrm{cm}$ to at least $10^{16}\,\mathrm{cm}$ with a mass-loss rate of 10$^{-4}$ $\rm M_{\odot}\ yr^{-1}$, both assuming a wind-velocity of 10 $\rm km\ s^{-1}$. The early nebular phase observations display an axisymmetric line profile of [\ion{O}{1}], red-ward attenuation of the emission of H$\rm \alpha$ post 125 days and flattening in the $Ks$-band marking the onset of dust formation.

\end{abstract}


\section{Introduction}
\label{sec:intro}

Massive stars ($>$\,8\,$M_\odot$, \citealp{2003heger}) reach the termination of their evolutionary phase upon exhaustion of their nuclear fuel and undergo core-collapse supernovae (CCSNe). SNe II form a subset of CCSNe that show Balmer features in their spectral sequence \citep{1997filippenko}. The progenitors of these SNe are confirmed to be red supergiants (RSGs) through pre-explosion imaging of several SNe II in nearby galaxies \citep{2009smartt, 2017vandyk}. 

Massive stars are thought to have steady stellar winds driven by radiation pressure \citep{1986chiosi}. The Galactic RSGs show an observed mass-loss rate in the range of $\rm 10^{-7} - 10^{-5} \ M_{\odot}\ yr^{-1}$ \citep{2018beasor}, due to its dependence on their luminosity and effective temperature. However, early spectral observations of SNe II have shown a short-lived `flash' phase with narrow high-ionization features, indicating a much higher mass loss rate than steady winds \citep{2022dessart}. Such features were first studied extensively in SN~1998S \citep{2000gerardy} as they trace the confined CSM arising from late stages of mass-loss in RSGs \citep{2017yaron}. A significant fraction of SNe II discovered by Zwicky Transient Facility (ZTF) exhibit such spectroscopic signatures of early interaction during their evolution ($\sim$\,36\%, \citealp{2023bruch}). 

The effects of a dense CSM driven by high mass loss rates have been known to delay shock breakout \citep{2018forster}. It occurs when the shock encounters a medium of significantly low optical depth, allowing the radiation to escape in front of the shock \citep{2017waxman}. The breakout is the first electromagnetic signature in CCSNe and happens within a few hours after core-collapse for a typical RSG \citep{1999matzner}. However, if the massive star is engulfed in a compact, dense CSM, the shock breakout leads to an elongated and brighter electromagnetic signature \citep{2010ofek,2011chevalier}, affecting the UV light curves and color evolution \citep{2021Nathiramatsu, 2022terreran, 2023irani}. The emergence of high-cadence all-sky surveys \citep{2018forster, 2023subrayan} coupled with prompt follow-up observations has photometrically revealed that fast-rising light curves are primarily reproduced via interaction with CSM. Such an interaction with CSM complicates the classification of SNe II \citep{2019aavinash}. SNe II are primarily classified into two main sub-types \citep{1979barbon} based on their plateau phase curve decline rates: a Type IIP SN with a slow decline ($\rm< 1.5\ mag\ (100\,d)^{-1}$) and a Type IIL SN with a steep linear decline ($\rm> 1.5\ mag\ (100\,d)^{-1}$) \citep{2016valenti}. The primary distinction between Type IIP/L SNe arose from the divide between the outer hydrogen envelope mass \citep{2003heger, 2018eldridge, 2021Hiramatsu}. However, studies by \citet{2017moriya,2018moriya,2017morozova} have suggested that a higher decline rate during the plateau phase is primarily driven by interaction with CSM, its compactness, and geometry \citep{2019andrews, 2022galan}. \citet{2021chugai} also indicated that higher CSM mass and wind density have been inferred for steeper declining SNe IIL compared to SNe IIP. 

Detailed studies of highly interacting SNe II such as SN~2013fs \citep{2017yaron}, SN~2020pni \citep{2022terreran}, SN~2020tlf \citep{2022galan} etc, have suggested a much higher mass-loss rate for their RSG progenitors, i.e., $\rm >\ 10^{-3}\ M_{\odot}\ yr^{-1}$. Since steady stellar winds cannot propel such a high mass loss rate in an RSG \citep{2008puls}, the paradigm has shifted towards a more complex scenario in recent years. These mechanisms include eruptive bursts of mass loss \citep{2006smith}, extensive RSG mass loss through surface instability \citep{2010yoon}, wave-driven mass loss during the end stages of nuclear burning \citep{2012quataert} and/or turbulent convection driven by dynamical instability \citep{2014smith}. The complex mass-loss mechanisms markedly influence the properties of a CCSN through the diverse pathways of forming CSM and its resulting geometrical configuration. 

Understanding the explosion geometry of the progenitor of SNe II and its CSM without directly resolving the source is aided by polarimetric observations. Polarimetric studies of CCSNe have indicated that polarization in CCSNe arises from Thomson scattering in the SN atmosphere \citep{1991jeffery, 2008wang}. Extensive studies by \citet{2001leonard, 2006leonard, 2019nagao} have highlighted numerous SNe II displaying low polarization during the photospheric phase and a subsequent increase in polarization during the transition to the nebular phase, suggesting the presence of an aspherical explosion. Some peculiar SNe II, such as SN~2017gmr \citep{2019nagao}, have exhibited an early onset of polarization during the photospheric phase, indicating an asymmetric CSM. The interactions in SNe II help probe the geometric configuration of the CSM, as the shock heating of the CSM primarily drives the early-phase luminosity. This allows for probing various configurations of the CSM, including asymmetries resembling a disk, torus, and asymmetric shell, and identifying clumpiness within the CSM \citep{2015smith}.

The geometry of the CSM helps constrain the physical mechanism of its formation from the progenitor and sheds light on the observed features. Inputs from the line profiles of narrow and intermediate-width features during the early phase also offer insights into the geometry of the CSM, in addition to its density structure \citep{2011asmith}. Imaging polarimetry provides a general overview of the large-scale structure of the explosion, whereas spectropolarimetry aids in investigating polarization across different elemental features to characterize their distribution \citep{2017branch}.

This leads us to the nearest SN II in the decade, SN~2023ixf, on which several studies have been published. SN~2023ixf was discovered by Koichi Itagaki \citep{2023ixfitagaki} at JD 2460084.4 and reported on Transient Name Server (TNS) followed by a prompt classification as a Type II SN \citep{2023perley} with visible signatures of flash ionization emission features. Early observational studies detailed the evidence of a shock breakout within a dense CSM \citep{2023yamanaka, 2023galan, 2023smith, 2023hosseinzadeh, 2023bostroem, 2023teja, 2023hiramatsu, 2024zimmerman, 2024yang, 2024li}. The evidence of asymmetry in the dense CSM was further discussed by \citep{2023smith, 2023vasylev, 2024li}. The SN was also detected in X-rays 4 days beyond explosion \citep{2023grefenstette, 2023chandra, 2024zimmerman} as well as in radio \citep{2023matthews} along with tight non-detections in mm-wavelengths \citep{2023berger} indicating strong signatures of interaction. Pre-explosion imaging studies \citep{2023kilpatrick, 2023pledger, 2023vandyk, 2024neustadt, 2023jencson, 2023soraisam, 2023niu, 2023qin, 2024ransome, 2024xiang} revealed the presence of an RSG progenitor embedded in large amounts of dust with indications of periodic variability. However, the inferred progenitor masses vary significantly, ranging from 8 to 24\ $\rm M_{\odot}$. Several other studies have established multi-wavelength limits on progenitor activity \citep{2023basu, 2023dong, 2023matsunaga, 2023kong, 2023panjkov}.

Our work presents the multi-wavelength analysis of SN~2023ixf through hydrodynamic light curve modeling until the early nebular phase and long-term spectroscopic and polarimetric monitoring, highlighting the signatures of asymmetries in SN~2023ixf. The data spanning the first 20 days of our study have already been published in \citet{2023teja} discussing the early photometric and spectroscopic analysis. We adopt the explosion epoch as estimated by \citet{2023teja}, i.e., JD 2460083.315, and the epoch described hereafter in the text is described with respect to this date. We provide an overview of the host galaxy's properties in Section~\ref{sec:host}, and summarize the details on observational data acquisition and reduction in Section~\ref{sec:obsdata}. The photometric properties of SN2023ixf are analyzed and discussed in Section~\ref{sec:photevol}, followed by a discussion on its spectroscopic evolution in Section~\ref{sec:specevol}. Section~\ref{sec:pol} delves into the temporal evolution of the polarimetric characteristics. Section~\ref{sec:modelmesa} describes the hydrodynamical light curve modeling to infer the properties of the progenitor and its multi-faceted CSM. We qualitatively discuss the presence of a confined CSM and the signatures of asphericity in the CSM/ejecta in order to understand the progenitor of SN~2023ixf in Section~\ref{sec:discussion} and summarize our results in Section~\ref{sec:summary}.


\section{Host Galaxy - M101}
\label{sec:host}

\begin{figure}[hbt!]
\centering
	 \resizebox{\hsize}{!}{\includegraphics{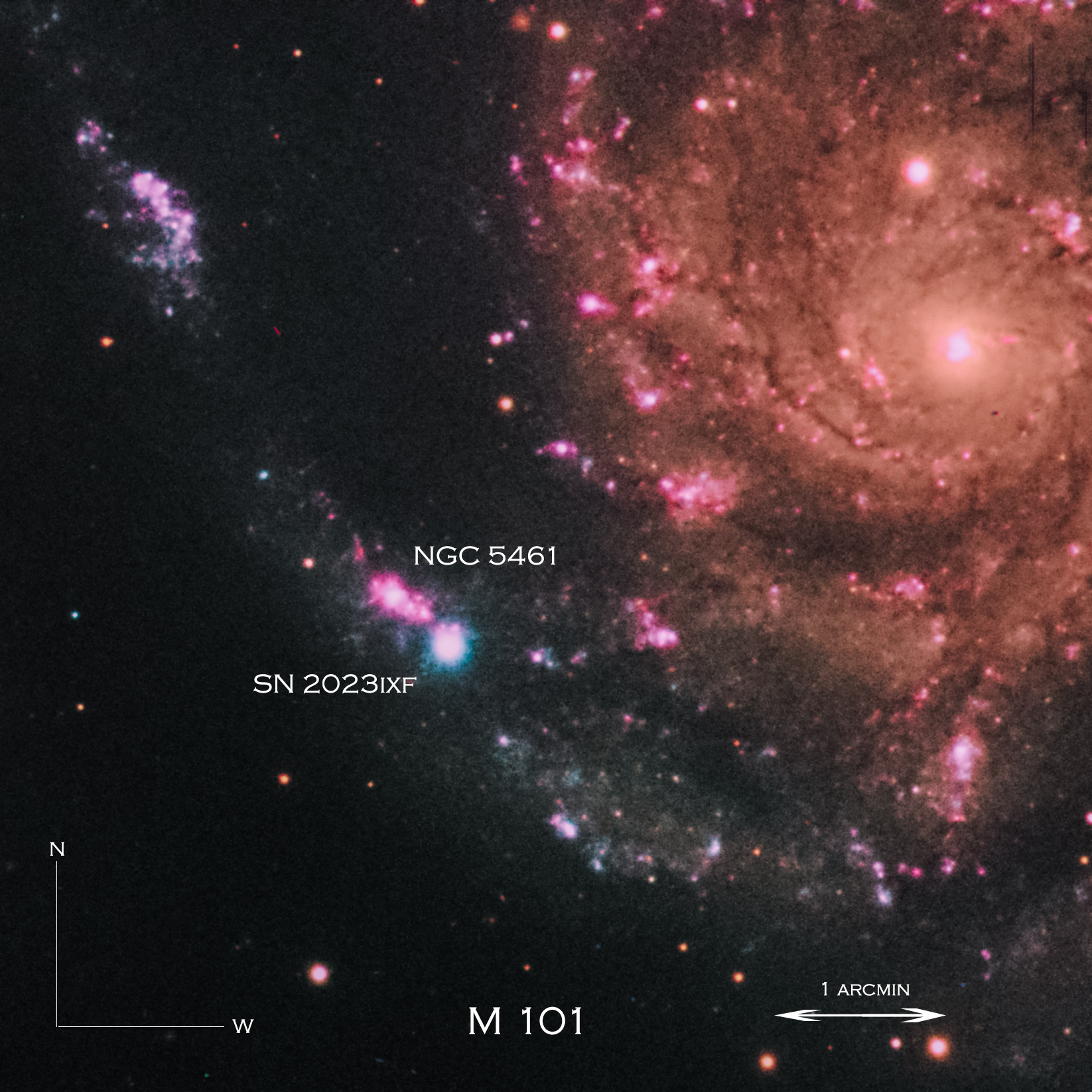}}
    \caption{color composite image (RGB) of the host galaxy M~101 using $r'$ (red), $g'$ (green), and $u'$ (blue) acquired from 2-m HCT. SN~2023ixf and the nearby Giant \ion{H}{2} region NGC~5461 are labeled.}
    \label{fig:m101}
\end{figure}

SN~2023ixf exploded in the outer spiral arm of the face-on spiral Galaxy, M~101 (NGC~5457), located in the constellation Ursa Major at a redshift of z\,=\,0.0008046 obtained from NASA Extragalactic Database (NED\footnote{\url{https://ned.ipac.caltech.edu}}). SN~2023ixf is the 3rd closest CCSN in this millennium after SN~2004dj \citep{2006vinko} and SN~2008bk \citep{2012vandyk}. SN~2023ixf lies proximal to one of the giant \ion{H}{2} regions in M~101, namely NGC~5461 (see Figure~\ref{fig:m101}), and in the immediate neighborhood of the \ion{H}{2} region \#1086 \citep{1990hodge}. Using emission-line diagnostics of the spectra of \ion{H}{2} region \#1086, \citep{2023vandyk} estimated an oxygen abundance through of 8.43\,$\rm \lesssim$\,12\,+\,log[O/H]\,$\rm \lesssim$\,8.86, which equates to a metallicity of 0.10\,$\rm \lesssim$\,Z\,$\rm \lesssim$0.20 close to the site of the SN. We adopt a mean distance of 6.82\,$\pm$\,0.14 Mpc ($\rm \mu$ = 29.17\,$\pm$\,0.04 mag, \citealp{2015trgb, 2022riess}), a Galactic reddening of $E(B-V)$\,=\,0.0077$\pm$0.0002 mag \citep{2011schlafly}, the host reddening of $E(B-V)$\,=\,0.031 $\pm$ 0.011 mag \citep{2023TNSAN.160....1L}, and a total reddening of $E(B-V)$\,=\,0.039\,$\pm$\,0.011 mag for SN~2023ixf.

\section{Data Acquisition and Reduction}
\label{sec:obsdata}

\subsection{Ultraviolet/Optical/Near-Infrared Photometry}

The optical photometric observations of SN~2023ixf were performed using the robotic 0.7-m GROWTH-India telescope (GIT, \citealp{2022growth}) located at the Indian Astronomical Observatory (IAO) in Hanle, India. The observations were carried out in the Sloan Digital Sky Survey (SDSS) $u'g'r'i'z'$ filters beginning 2023 May 20 UT. The data were processed with the standard GIT image processing pipeline described in \citet{2022growth}, and the steps followed are described in \citet{2023teja}. We also carried out photometric observations in $BVRI$ using a 0.36-m Schmidt Cassegrain telescope (Celestron EdgeHD 1400) at the Home observatory in Nayoro, Hokkaido, utilizing a CCD FLI ML1001E camera with an IDAS filter (standard system). The data reduction and aperture photometry were carried out using the software MIRA Pro x64 \citep{MiraProx64}. The calibration was performed using the stars from APASS catalog \citep{2013zacharias}. Additionally, optical photometric observations in $VRI$ bands were also carried out using Atik 460 EX Mono CCD mounted on the 0.61-m Vasistha telescope at Ionospheric and Earthquake Research Centre and Optical Observatory (IERCOO), Sitapur, ICSP, Kolkata. Photometric calibrations were done using Tycho software and the ATLAS catalog \citep{2018tonry}. $BVRI$-band imaging was also carried out using the 0.51-m telescope at Oku Observatory, Okayama, using the SBIG Camera STXL-6303. The data reduction was performed according to the standard procedure using \texttt{IRAF}.

We monitored SN~2023ixf in the near-infrared (NIR) from the Hiroshima Optical and Near-InfraRed Camera \citep[HONIR;][]{2014akitaya} mounted on the 1.5-m Kanata Telescope located at Higashi-Hiroshima Observatory, Hiroshima University, Japan. Near-infrared observations were also carried out using kSIRIUS\footnote{The design of kSIRIUS is adopted from SIRIUS (the near-infrared simultaneous three-band camera}, the near-infrared simultaneous $JHKs$-band imager attached to the Cassegrain focus of the 1.0-m telescope at the Iriki Observatory in Kagoshima, Japan. The NIR data were reduced using standard procedures in \texttt{IRAF} (Image Reduction and Analysis Facility\footnote{\url{https://github.com/iraf-community/iraf}}), and the photometric magnitudes were obtained through the point-spread-function (PSF) photometry using standard \texttt{IRAF} tasks such as DAOPHOT \citep{Stetson1987}. The photometric calibration was performed using secondary stars from the 2MASS catalog \citep{2006AJ....131.1163S}.

The Ultraviolet Optical Telescope (UVOT; \citealp{2005roming}) onboard the Neil Gehrels \textit{Swift} Observatory \citep{2004gehrels} monitored SN~2023ixf extensively beginning May 21, 2023. We reduced the publicly available data obtained from Swift Archives\footnote{\href{https://www.swift.ac.uk/swift_portal/}{Swift Archive Download Portal}}. Photometry was performed using the UVOT data analysis software in \texttt{HEASoft}, following the procedure described in \citet{2022teja}. Upon using a 5$\arcsec$ aperture, a significant part of the nearby \ion{H}{2} contaminates the flux, especially in the ultraviolet (UV) bands at later epochs ($>$ 20\,d). We used the archival data for the host M~101 obtained on Aug 29, 2006, with OBSID 00035892001, available in the \textit{Swift} archive, as template images for removing the host contribution. The flux obtained at the SN site in the template images is comparable to the late-phase fluxes in all the swift bands. We employed \texttt{Swift\_host\_subtraction}\footnote{\href{https://github.com/gterreran/Swift_host_subtraction}{https://github.com/gterreran/Swift\_host\_subtraction}} code \citep{2009brown, 2014brown} to remove the host contribution.

We supplemented our multi-wavelength light curve data with early phase photometry ($<$\,10\,d) in $griz$ from \citet{2023galan}. We also utilized the streak photometry performed on the saturated UV bands ($UVW2, UVM2, UVW1$) from \citet{2024zimmerman}. 

\subsection{Optical Spectroscopy}

Low-resolution optical spectroscopic observations of SN~2023ixf were carried out using the Himalayan Faint Object Spectrograph (HFOSC) instrument mounted on the 2-m Himalayan Chandra Telescope (HCT) at IAO \citep{2014Prabhu}. The HFOSC observations were performed in grisms Gr\#7 (3500-7800 \AA, R\,$\sim$\,500) and Gr\#8 (5200-9250 \AA, R\,$\sim$\,800). The spectroscopic data were reduced in a standard manner using the packages and tasks in \texttt{IRAF} with the aid of the Python scripts hosted at \textsc{RedPipe} \citep{2021redpipe}. Optical spectroscopic observations were also performed using the Kyoto Okayama Optical Low-dispersion Spectrograph with optical-fiber Integral Field Unit (KOOLS-IFU, \citealp{2019matsubayashi}) mounted at the 3.8-m Seimei Telescope \citep{2020kurita} located in Okayama Observatory, Kyoto University, Japan. The KOOLS-IFU observations were carried out using VPH-blue (4100-8900 \AA, R\,$\sim$\,500). The data reduction was performed using the Hydra package in \texttt{IRAF} and a reduction software developed for KOOLS-IFU data\footnote{\url{http://www.o.kwasan.kyoto-u.ac.jp/inst/p-kools}}. Arc lamps of Hg, Ne, and Xe were used for wavelength calibration. We obtained additional optical spectroscopic data using the TriColor CMOS Camera and Spectrograph (TriCCS) installed on the 3.8-m Seimei telescope having a wavelength coverage of 4000–10500 \AA\ (R\,$\sim$\,700) with grisms g, r and iz. We used the L. A. Cosmic pipeline \citep{2001vandokkum} to remove cosmic rays. We used arc lamps (Hg, Ne, and Xe) for wavelength calibration, and the flux was calibrated using spectrophotometric standards. All the spectra have been continuum calibrated with respect to $gri$ photometry and corrected for the redshift of the host galaxy.

\subsection{Imaging Polarimetry and Spectropolarimetry}

We carried out imaging polarimetric observations of SN~2023ixf using HONIR in {\it R}-band ($\lambda_{eff}$\,=\,0.65\,$\mu$m). The observations comprised a sequence of four position angles of the achromatic Half-Wave Plate (i.e., 0$^\circ$, 45$^\circ$, 22.5$^\circ$ and 67.5$^\circ$). The instrumental polarization is about 0.1\%, based on the observations of the unpolarized standard star HD 14069, which is consistent with past observations from HONIR \citep{2023imazawa}. The polarization angle was corrected by observing several polarized standards BD +59\textdegree 389, BD +64\textdegree 106 over different nights.

We obtained spectropolarimetric observations of SN~2023ixf over five epochs on 2023 June 14, 15, 16, 18, and 19 using the CCD Imaging/Spectropolarimeter (SPOL; \citealp{1992schmidt}) mounted on the 2.3 m Bok Telescope at Steward Observatory, Kitt Peak, Arizona. We used the 600 l/mm grating with a spectral coverage of 4000-7550\AA\ and obtained a spectral resolution of 26.53\,\AA\ using a slit size of 0.4 mm (corresponding to 4.1" in the sky). In addition to this, we used a Hoya L38 blocking filter with a cut-off wavelength of 380nm. We used polarized standards (Hiltner 960; HD 155528) to calibrate the polarization angle, with the average discrepancy between the observed and expected polarization angles for multiple standards being less than 1$\rm ^\circ$ \citep{1992schmidt}. Furthermore, the unpolarized standard (BD+28\textdegree4211) observations confirmed that SPOL exhibited low instrumental polarization, measuring at less than 0.1\%. We use the same data acquisition and reduction procedure mentioned in \cite{bilinski2018}. The polarized spectra corresponding to four exposures that sample 16 waveplate positions were flat-field corrected, background subtracted, and flux calibrated using the SPOL polarization data reduction pipeline to obtain the Stokes parameters q, u, p, and polarization angle.

\begin{figure*}
\centering
	 \resizebox{\hsize}{!}{\includegraphics{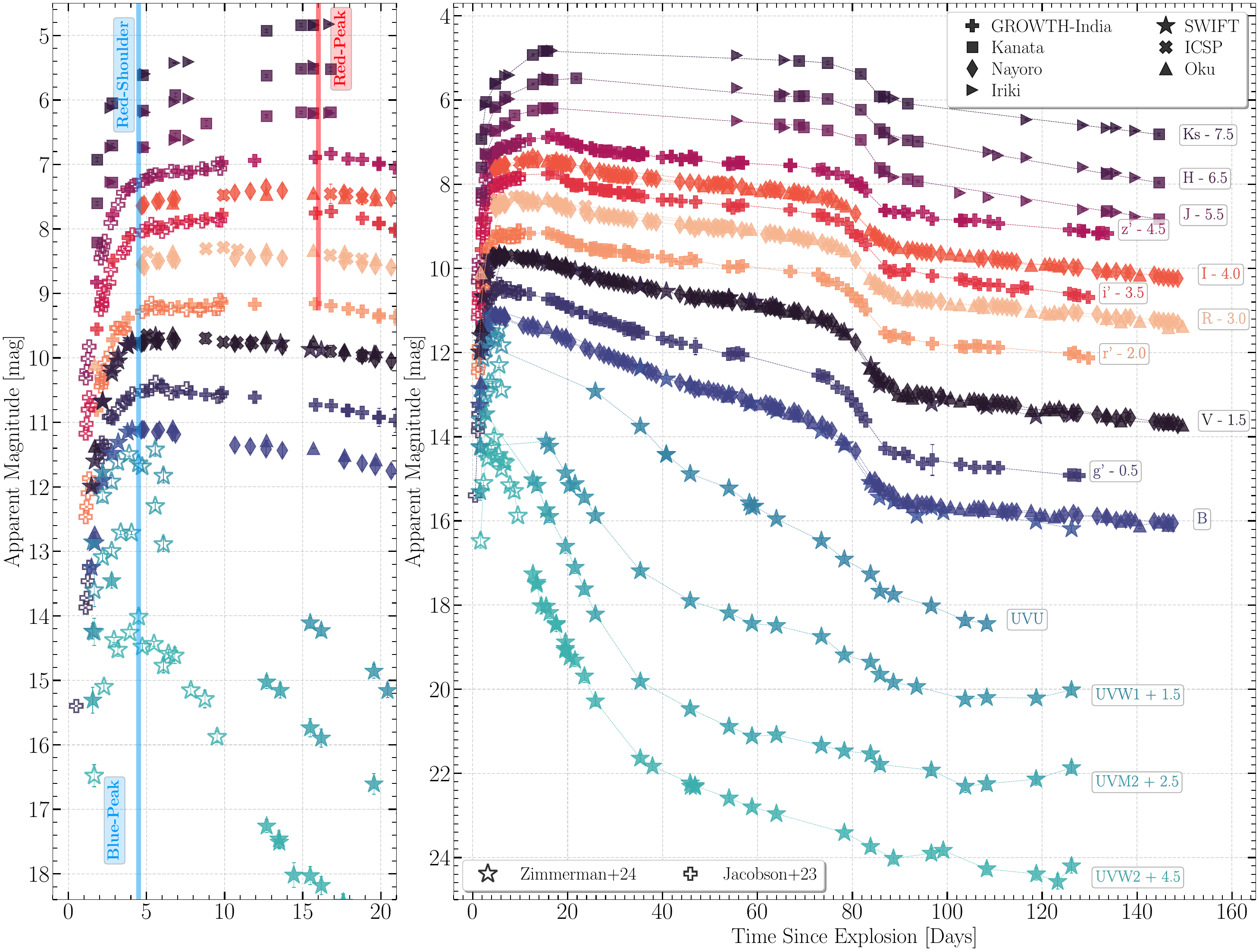}}
    \caption{Multi-wavelength photometry of SN~2023ixf spanning ultraviolet, optical, and near-infrared wavelengths. Template subtraction was performed only in the UV bands ($UVW2$, $UVM2$, and $UVW1$) due to non-negligible contamination. The left panel provides a close-up of the initial rise across all bands, annotated as \textit{blue peak \& red shoulder}, and the subsequent \textit{red peak}. (The photometric data is available as data behind the figure.)}
    \label{fig:applc}
\end{figure*}

\section{Photometric Evolution}
\label{sec:photevol}

Panchromatic light curves covering ultraviolet, optical, and near-infrared wavelengths from various facilities are shown in Figure~\ref{fig:applc}, displaying the well-sampled SN evolution spanning 1.4 d -- 150 d after the explosion. All the magnitudes presented in this paper were calibrated to the AB system using the transformations described in \citet{2007AJ....133..734B}.

\subsection{Rise times, decline rates and plateau-length}
\label{sec:risetime}

The light curve peaks were measured by fitting the light curves with a smooth spline, calculating the gradient ($\rm dm/dt$) of the fit, and identifying the zero-crossing points to determine the peaks. SN~2023ixf shows a sharp rise to the peak in the ultraviolet bands - $UVW2$ and $UVW1$ (i.e., \textit{blue peak}) with a rise time of $\sim$\,4.5\,d. The rise time is $\sim$\,6\,d in $V$-band, which is faster than the prototypical rise time of 10\,d in Type II SN~1999em \citep{2002bleonard}. The steeper rise to the maximum for SN~2023ixf is attributable to the interaction of the SN ejecta with CSM \citet{2018forster}. The light curves in the redder bands ($r'i'z'$) seem to show a steep rise to a distinct shoulder, i.e., \textit{red shoulder} post $\sim$\,5\,d and a gradual ascent to the maximum, i.e., \textit{red peak} at $\sim$\,16\,d. The \textit{red shoulder} is seen as an abrupt change of slope in the gradient of the early light curve. The presence of the \textit{red shoulder} is driven by the flux excess due to the interaction of SN~2023ixf with a confined CSM since it occurs immediately after the peak in the UV wavelengths (i.e., \textit{blue peak}). A similar shoulder and peak are seen in the simulated red-band light curves of SNe II with strong wind models ($>$ 10$^{-3}$ $\rm M_{\odot}\ yr^{-1}$) by \citet{2017dessart}, affirming an alternate scenario to look for signatures of early interaction in the light curves besides the rise time. The rise time for the \textit{red peak} in SN~2023ixf s higher than that of SNe IIP (7.0\,$\pm$\,0.3 d) and in the ballpark of SNe IIL (13.3\,$\pm$\,0.6 d) \citep{2015gall}.

Post-maximum, the multi-band light curves of SN~2023ixf settle onto a plateau of roughly $\sim$\,75\,d in $V$ and other redder bands before transitioning to the radioactively-powered tail phase at $\sim$\,90\,d. The plateau length is at the shorter end of the typical plateau length of 100~d for SNe II, hence putting SN~2023ixf amongst some of the seldom observed short plateau SNe \citep{2021Hiramatsu, 2022teja}. We estimated the plateau decline rates of SN~2023ixf following the prescription of \citet{2014anderson}. The $V$-band light curve of SN~2023ixf showed an early plateau decline rate (s1) of $2.70^{+0.48}_{-0.49}$ $\rm mag\,(100\,d)^{-1}$ and a late-plateau decline rate (s2) of $1.85^{+0.13}_{-0.14}$ $\rm mag\,(100\,d)^{-1}$. The late-plateau decline rate of SN~2023ixf is higher than the mean decline rate of 1.3 $\rm mag\,(100\,d)^{-1}$ inferred for SNe II \citep{2014anderson}. The tail phase decline rate (s3) of SN~2023ixf is $1.33^{+0.09}_{-0.09}$ $\rm mag\,(100\,d)^{-1}$ which is faster than the characteristic decline rate of $\rm^{56}$Co to $\rm^{56}$Fe (i.e., 0.98 $\rm mag\,(100\,d)^{-1}$) indicating incomplete trapping of $e^{+}$ and $\gamma$-rays.

\subsection{Light curve comparisons with other SNe II}
\label{sec:compsne}

\begin{figure}
\centering
	 \resizebox{\hsize}{!}{\includegraphics{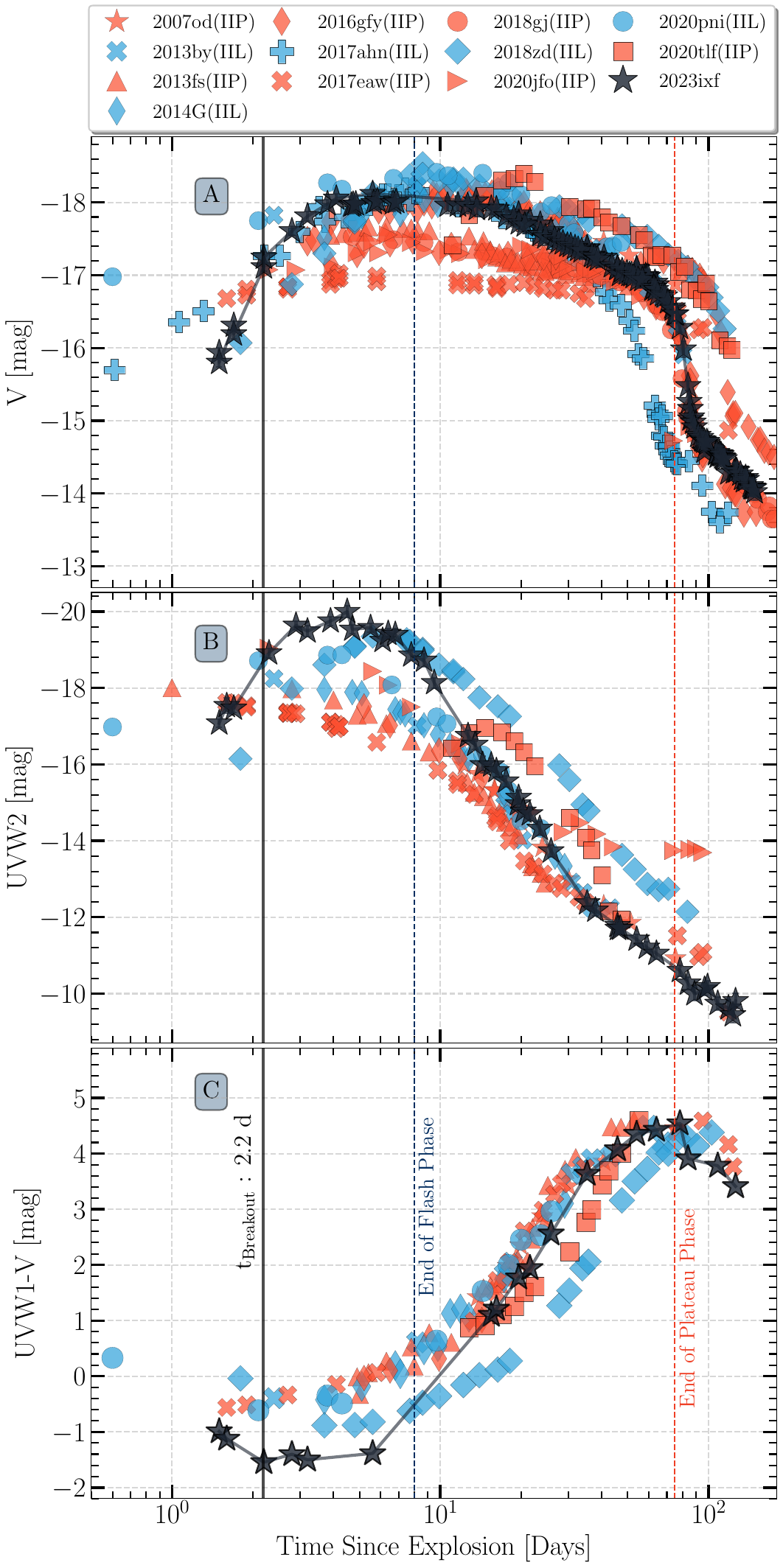}}
    \caption{Comparison of absolute magnitude light curves (in AB-mag) of SN~2023ixf in $UVW2$ and $V$ bands along with the $UVW1-V$ color evolution compared to other SNe II showing CSM interaction signatures.}
    \label{fig:compuv}
\end{figure}

The comparison of $V$-band absolute magnitude light curve of SN~2023ixf is shown in the top panel (A) of Figure~\ref{fig:compuv}. Owing to the short plateau length and fast declining nature of SN~2023ixf, we derived a comparison sample consisting of normal Type IIP SNe: SN~2007od \citep{2010andrews}, SN~2013fs \citep{2018bullivant}, SN~2016gfy \citep{2019aavinash}, 2017eaw \citep{2019szalai} and SN~2020tlf \citep{2022galan}; short-plateau Type IIP SNe: SN~2018gj \citep{2023teja18gj}, SN~2020jfo \citep{2022teja}, and Type IIL SNe: SN~2013by \citep{2015valenti}, SN~2014G \citep{2016terreran}, SN~2017ahn \citep{2021tartaglia}, SN~2018zd \citep{2021Nathiramatsu} and SN~2020pni \citep{2022terreran}. The peak $V$-band luminosity of SN~2023ixf is --18.2 mag, brighter than the average peak luminosity of SNe II (i.e., --16.74 $\pm$ 1.01 mag) inferred by \citet{2014anderson}, which is dominated by the population of slow-declining SNe II. The peak luminosity of SN~2023ixf is similar to the SNe IIL, namely SN~2013by, SN~2014G, and SN~2017ahn, but slightly fainter than that of SN~2020pni; however, it is brighter than the majority of normal and short-plateau SNe IIP. The plateau-decline rate of SN~2023ixf is similar to that of SN~2014G ($\sim$\,1.7 $\rm mag\,(100\,d)^{-1}$) and SN~2013by ($\sim$\,1.5 $\rm mag\,(100\,d)^{-1}$). The $\sim$\,75\,d plateau length of SN~2023ixf is similar to that of SN~2014G, SN~2013by and SN~2013fs. Overall, SN~2023ixf shows remarkable photometric resemblance in peak-luminosity, plateau decline rates, plateau length, and plateau drop to SN~2013by and SN~2014G.

We further compare SN~2023ixf to SNe II with CSM interaction, followed up extensively in UV by \textit{Swift} over the last two decades. The data was downloaded from the Swift Optical/Ultraviolet Supernova Archive (SOUSA\footnote{\url{https://pbrown801.github.io/SOUSA/}}, \citealp{2014sousa}) and the Vega-mag were transformed to AB-mag for consistency. In the exhaustive sample of SNe II observed by \textit{Swift}, the $UVW2$ and $UVW1$ bands exhibit a rapid surge in the flux, brightening by over $\sim$\,3 mags in 3 days, before reaching a peak magnitude of $\sim$ --20 mag at $\sim$\,4.5\,d. This prolonged brightening observed in the initial phase of the UV light curve indicates a shock breakout within a compact and dense CSM, leading to a more luminous and elongated shock breakout event \citep{2010ofek}. Such a distinct signature in the early UV light curves has been observed in only a limited number of SNe II. The detection of a UV burst extending over 1-day in PS1-13arp by \citet{2015gezari} was the first observation hinting at the possibility of the shock breaking out into a confined CSM. In our comparison sample, only SN~2018zd \citep{2021Nathiramatsu}, SN~2020pni \citep{2022terreran}, and SN~2020tlf \citep{2022galan} showed a similar intensification of the early phase UV light curve.

\begin{figure}
\centering
	 \resizebox{\hsize}{!}{\includegraphics{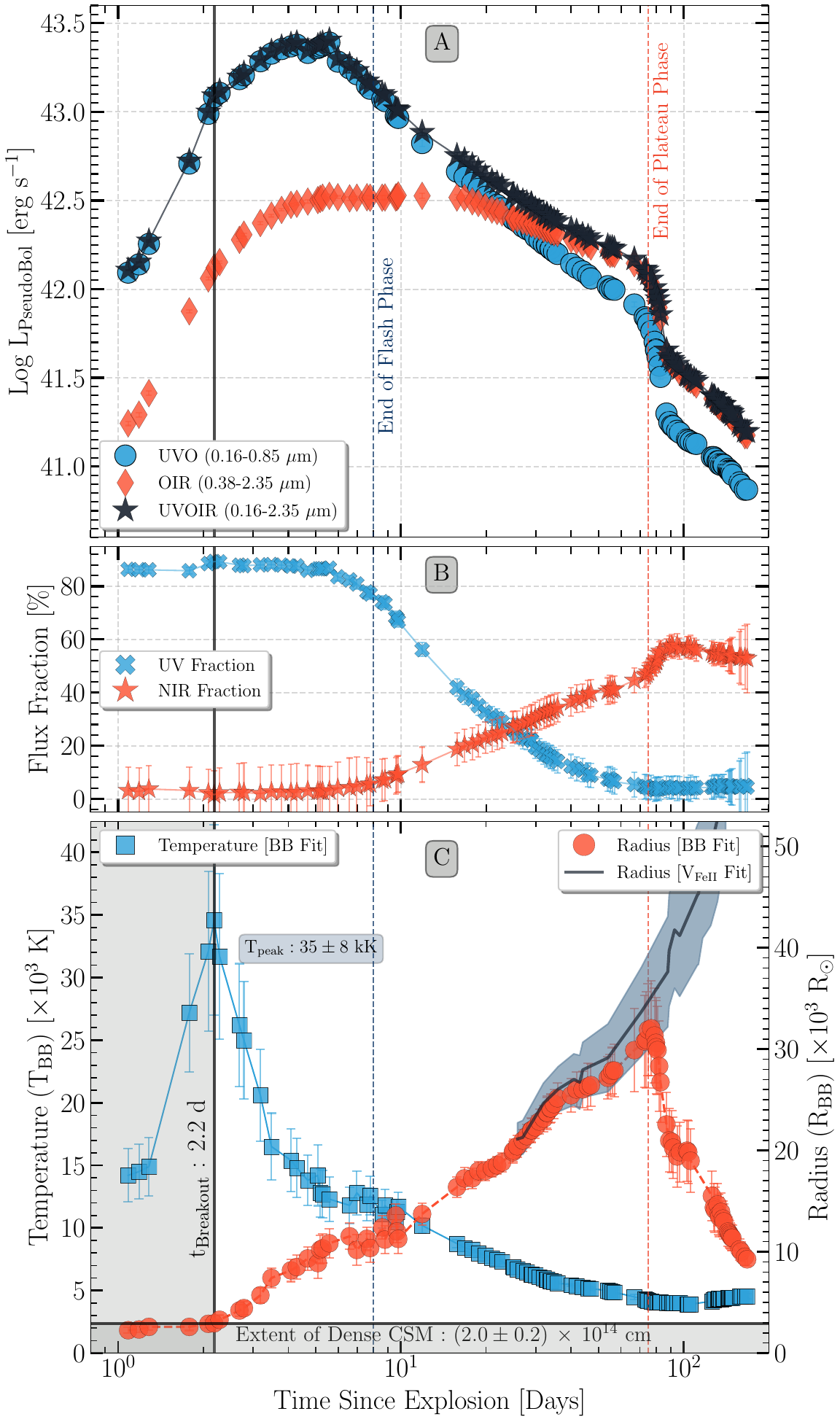}}
    \caption{\textit{Panel A}: Pseudo-bolometric light curves of SN~2023ixf computed in multiple wavelength bins. \textit{Panel B}: Temporal evolution of UV and NIR flux fraction in SN~2023ixf. \textit{Panel C}: Temperature and radius evolution of SN~2023ixf estimated from blackbody fits to the UVOIR data (0.16\,--\,2.35\,$\mu$m). The radius of the line-forming region, i.e., the photospheric radius estimated from \ion{Fe}{2}, is overplotted.}
    \label{fig:bollc}
\end{figure}

The evidence of the sharply declining UV flux was seen in the FUV spectra of SN~2023ixf obtained by \citet{2023teja} at $\sim$\,23\,d, which shows a featureless spectrum owing to the line-blanketing from the iron-group elements \citep{2009bufano}. Despite this, we still infer a decent contribution ($\sim$\,10\%, see Panel B in Figure~\ref{fig:bollc}) of UV flux to the overall bolometric light curve (i.e., a UV excess) during the mid-plateau phase ($\sim$\,38\, d). Our observations indicate continued interaction with CSM despite the absence of discernible signatures in the spectral sequence outlined in Section~\ref{sec:specphot}. This aligns with the inference from the theoretical models of interacting SNe II by \citet{2022dessart} that display early interaction frequently display a surplus of UV radiation during late phases.

\subsection{Bolometric Light Curve, Temperature and Radius Evolution}
\label{sec:boltemp}

We computed the pseudo-bolometric light curves of SN~2023ixf using \texttt{SuperBol} \citep{Nicholl2018}. The bolometric luminosity, color temperature ($\rm T_{col}$), and radius ($\rm R_{BB}$) evolution of the layer of thermalization were estimated using the blackbody fits to the SED of the SN at each epoch. The missing data in certain filters over the intermediate epochs was interpolated using a low-order spline. We computed the bolometric light curve in 3 wavelength bins, i.e., UVOIR (0.16\,--\,2.35\,$\mu$m), UVO (0.16\,--\,0.85\,$\mu$m) and OIR (0.38\,--\,2.35\,$\mu$m) and is shown in Panel~A of Figure~\ref{fig:bollc}. For the integration, the blue boundary for the UV and optical wavelengths is defined by the blue edge of the Swift $UVW2$ band and the $B$ band, respectively. The red boundary for the optical and infrared wavelengths is defined by the red edge of the $I$ band and the $Ks$ band, respectively. Henceforth, we refer to the UVOIR pseudo-bolometric light curve as the bolometric light curve. 

As discussed by \citet{2024zimmerman}, the color temperature and radius evolution shown in Panel~C of Figure~\ref{fig:bollc}, shows a steep increase in the temperature from around 14,000 K to 35,000 K over a duration of 2.2\,$\pm$\,0.1\,d and a relatively-flat radius evolution at (2.0\,$\pm$\,0.2)\,$\times$\,$\rm 10^{14}$\,cm (13\,$\pm$\,1 AU). The extended heating at a near-flat radius, in addition to the prolonged brightening in the UV flux, as discussed in Section~\ref{sec:compsne} are signs of further heating upon interaction with the dense CSM. From the photometric observations at 1.1\,d, the bolometric luminosity, $L_{BOL}$ = (1.28\,$\pm$0.11)\,$\times$\,$\rm 10^{42}\ erg\ s^{-1}$, and effective temperature of (14\,$\pm$\,2)\,$\times$\,$\rm 10^{3}\ K$ rises ten folds and two folds, respectively, in a span of just $\sim$\,1\,d. However, due to the ensued heating, the bolometric light curve of SN~2023ixf flux peaked later on $\sim$\,4.5\,d with a luminosity of 2.5\,$\pm$\,0.3\,$\times$\,$\rm10^{43}\ erg\ s^{-1}$. 

The temporal evolution of UV and NIR fraction of the pseudo-bolometric luminosity is shown in Panel~B of Figure~\ref{fig:bollc}. The contribution of the UV flux (0.16\,--\,0.38\,$\mu$m) to the bolometric flux stays at roughly about 85\% even until the bolometric maximum (same as the UV peak). If we ignore the early UV data for SN~2023ixf, the OIR bolometric light curve underestimates the bolometric luminosity by an order of magnitude, emphasizing the importance of UV observations of infant CCSNe and its importance to detailed hydrodynamical modeling. During the early nebular phase, the NIR flux contributes almost 60\% to the bolometric luminosity whereas the UV flux contributes roughly 5\%, highlighting the importance of NIR observations in determining $^{56}$-Ni mass in SNe II.

\subsection{\texorpdfstring{$^{56}$}~Ni Mass}
\label{sec:56Ni}

\begin{figure*}[!hbt]
\centering
	 \resizebox{0.97\hsize}{!}{\includegraphics{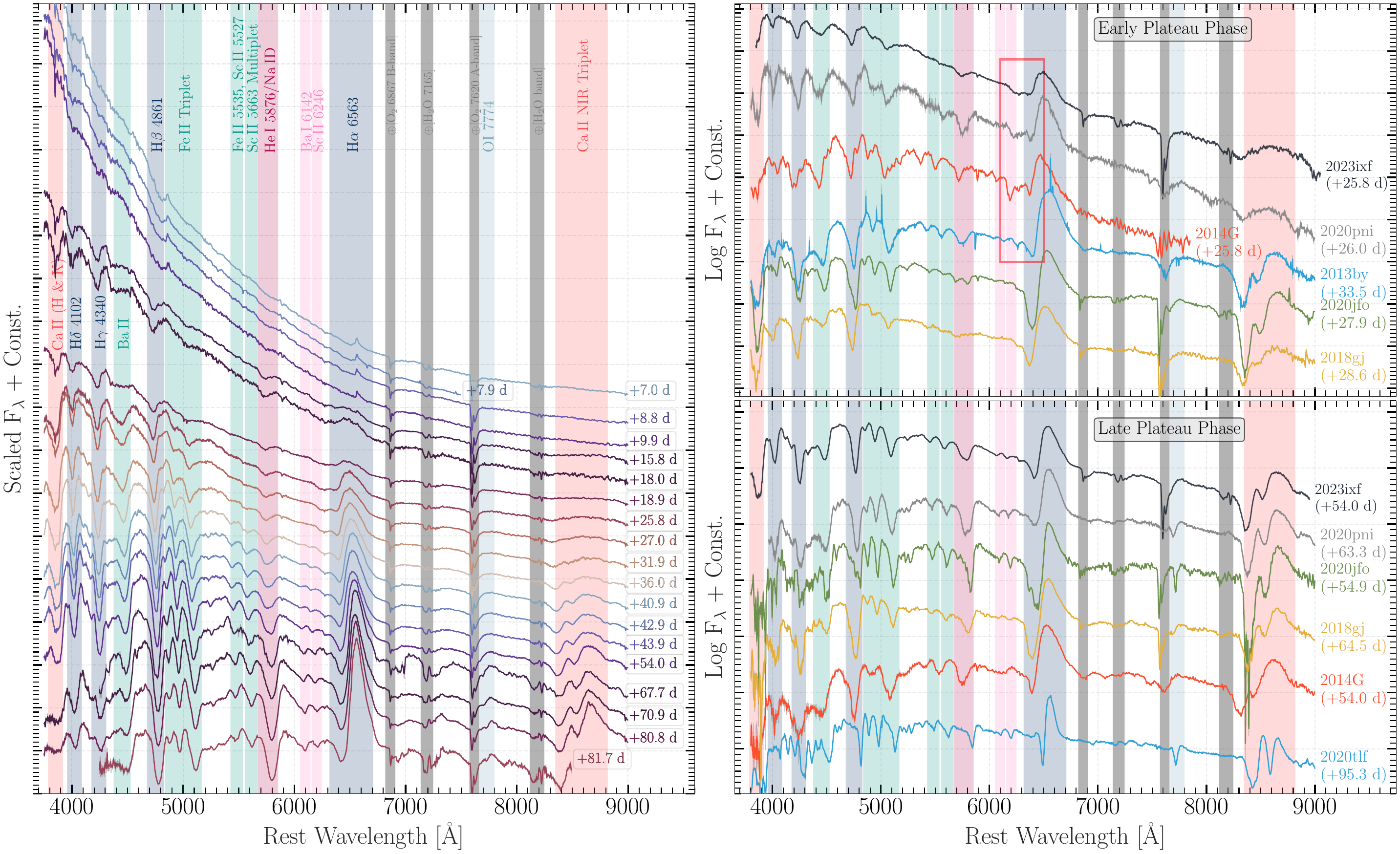}}
    \caption{\texttt{Left Panel:} Photospheric phase evolution of SN~2023ixf. Prominent spectral features are labeled and depicted using colored shaded areas. \texttt{Right Panels:} Comparison of the early and late plateau phase spectrum of SN~2023ixf with other SNe II from the literature with signatures of CSM interaction in the top and bottom panels, respectively. References: SN~2013by \citep{2015valenti}; SN~2014G \citep{2016terreran}; SN~2018gj \citep{2023teja18gj}; SN~2020jfo \citep{2022teja}; SN~2020pni \citep{2022terreran}; SN~2020tlf \citep{2022galan}. (The spectroscopic data is available as data behind the figure.)}
    \label{fig:specp}
\end{figure*}

Upon recombination of all the hydrogen in the outer envelope, SNe II post the plateau phase transitions to the tail phase powered by the decay of $\rm ^{56}Ni \rightarrow\ ^{56}Co \rightarrow\ ^{56}Fe$, which thermalizes the SN ejecta through the emission of $\gamma$-rays and $e^{+}$. We computed the $\rm ^{56}$Ni-mass using the relation postulated by \citet{2003hamuy}. The mean tail luminosity of SN~2023ixf around $\sim$\,145 d yields a $\rm ^{56}Ni$ mass of $\rm 0.054\pm0.006\ M_{\odot}$. Upon comparison with SN~1987A at a similar phase which synthesized a $\rm ^{56}Ni$-mass of 0.075\,$\pm$\,0.005 $\rm M_{\odot}$ \citep{1998turatto}, a $\rm ^{56}Ni$-mass of $\rm 0.054\pm 0.005\ M_{\odot}$ is estimated for SN~2023ixf. The above techniques assumed complete trapping of $\gamma$-rays by the SN ejecta. However, that is not always true, especially for short-plateau SNe due to a thinner hydrogen envelope \citep{2022teja} and may lead to underestimation of the $\rm ^{56}Ni$ yield. Hence, we modeled the late-phase UVOIR bolometric light curve of SN~2023ixf spanning 100 -- 150\,d using the analytical formulation from \citet{2016yuan}. We derived a total $\rm^{56}Ni$-mass of 0.059\,$\pm$\,0.001 $\rm M_{\odot}$ and a characteristic $\gamma$-ray trapping timescale of $\sim$\,220\,$\pm$\,3\,d, indicating a short-lived trapping and a shallower envelope. It is important to note that our parameter estimates are based on data spanning up to 150 days, which may result in a slight overestimation of the $^{56}$Ni mass. Therefore, we consider this an upper limit, as there are likely additional contributions to the light curve during the tail phase from CSM interaction, as discussed in Section~\ref{sec:compsne}.

\section{Spectroscopic Evolution}
\label{sec:specevol}

\subsection{Flash Spectroscopy Phase \texorpdfstring{$\lesssim$}~ 8 d}
\label{sec:specflash}

Earlier spectroscopic studies on SN~2023ixf \citep{2023yamanaka, 2023galan, 2023smith, 2023bostroem, 2023teja, 2023hiramatsu, 2024zimmerman, 2023zhang} showed a prominently blue continuum coupled with highly ionized features of \ion{C}{4}, \ion{N}{4}, \ion{C}{3}, \ion{N}{3} and \ion{He}{2} along with Balmer features beginning 1.1\,d from the estimated date of explosion. The features suggest flash-ionization and recombination of the confined dense CSM engulfing the progenitor of SN~2023ixf. Notably, these studies observed an increase in the relative line strengths of the highly ionized species such as \ion{N}{4} and \ion{C}{4} in contrast to \ion{N}{3} and \ion{C}{3}, respectively, implying a delayed ionization of the CSM. The increase in line fluxes of \ion{N}{4}, \ion{C}{4} and \ion{He}{2} noticed in the spectral sequence of SN~2023ixf spanned $\sim$\,4\,d from the explosion \citep{2023smith}, which roughly corresponds to the peak of UV light curves. This delayed ionization is in concurrence with the inference from the increase in temperature evolution in Section~\ref{sec:boltemp}, as it affirms the breakout of the shock from a confined dense CSM. The \ion{He}{2} $\rm \lambda$ 4686 vanished in the high-cadence spectroscopic observations presented in \citet{2023bostroem} between 7.6\,d to 8.4\,d indicating that the flash spectroscopy phase lasted approximately 8 days. 

\subsection{Photospheric Phase: 8 - 80 d}
\label{sec:specphot}


\begin{figure*}[hbt!]
\centering
	 \resizebox{0.22\hsize}{!}{\includegraphics{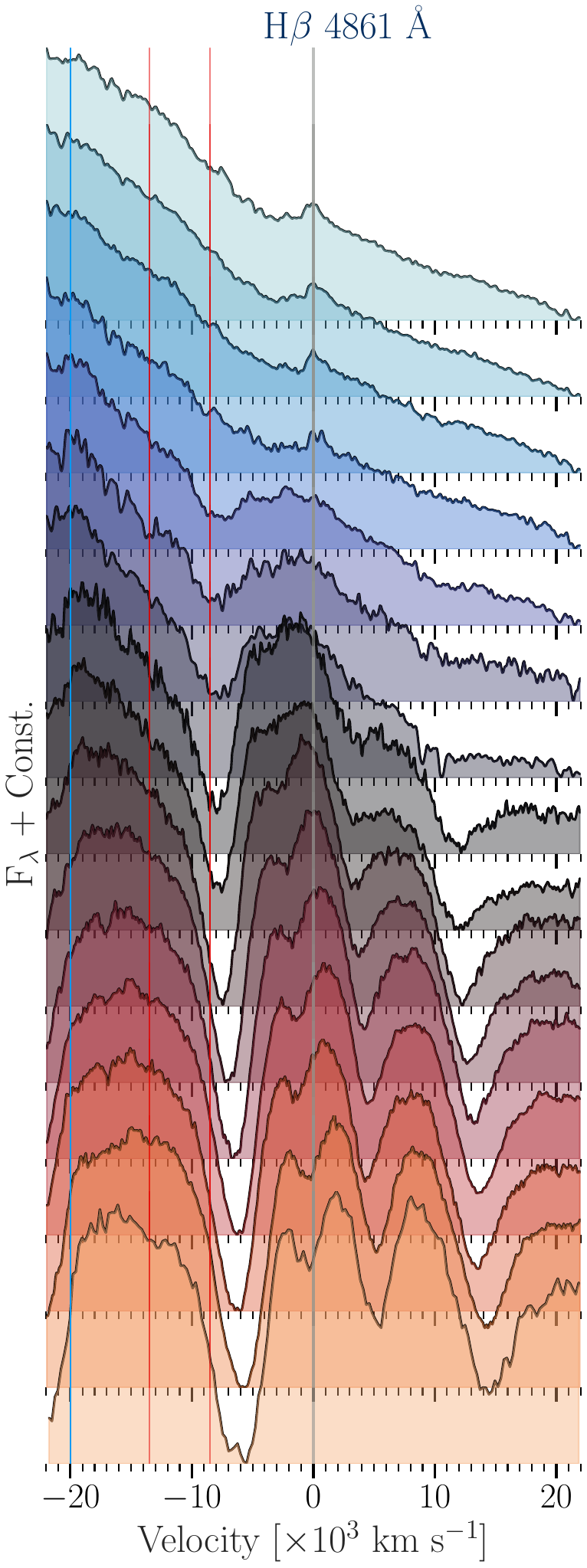}}
  \resizebox{0.26\hsize}{!}{\includegraphics{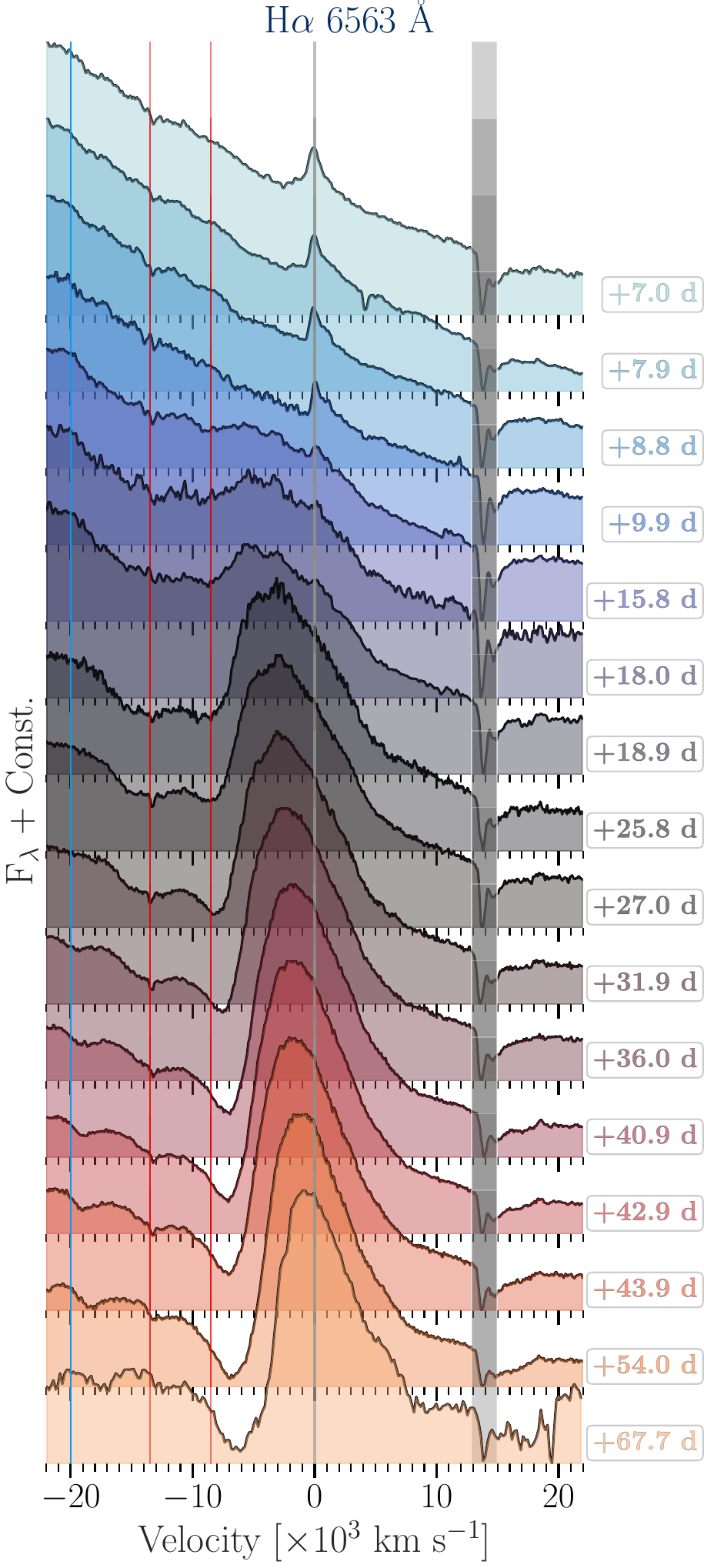}}
  \resizebox{0.45\hsize}{!}{\includegraphics{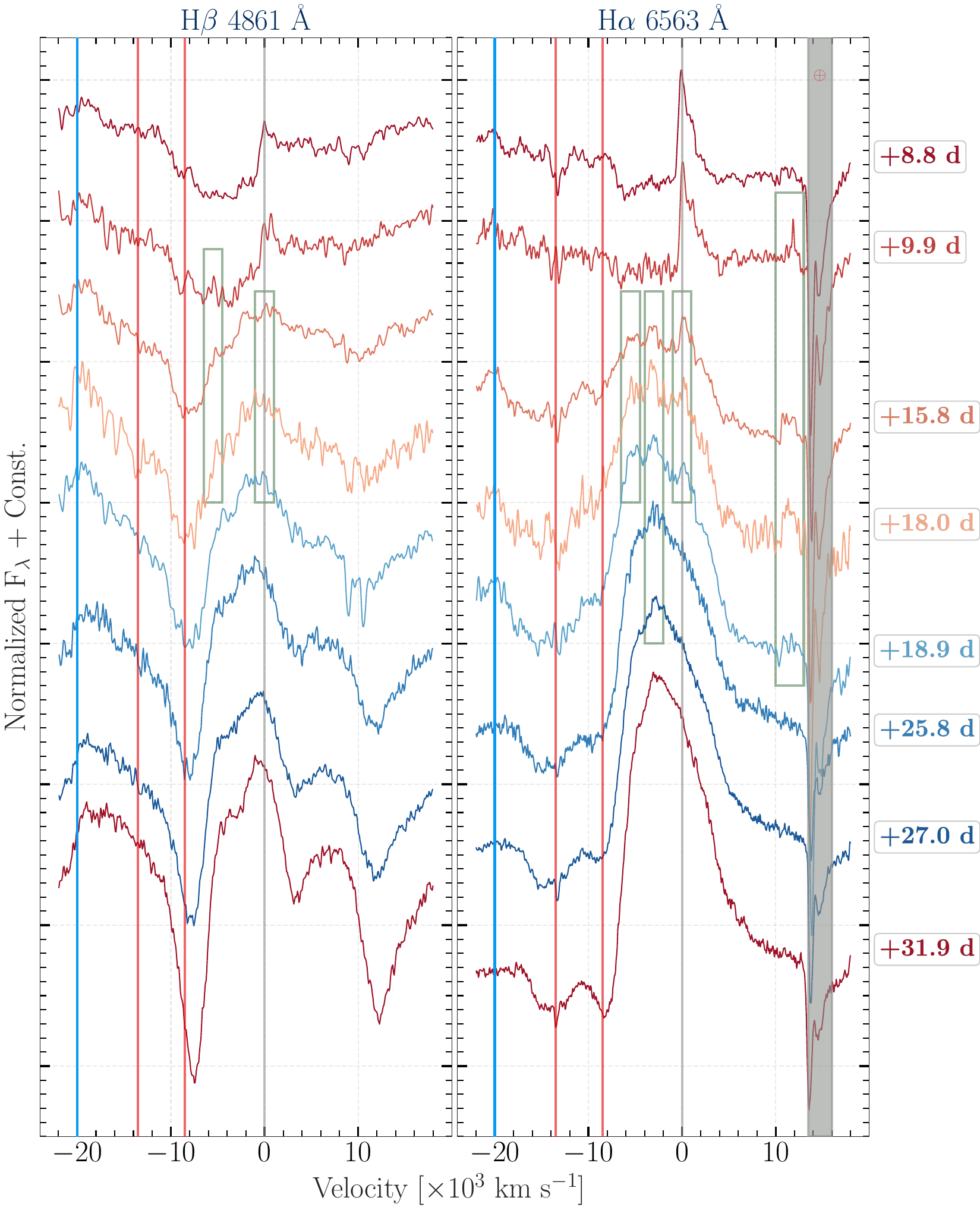}}

    \caption{Temporal evolution of $\rm H\beta$ and $\rm H\alpha$ profile of SN~2023ixf from the decline of the flash-ionization phase until the late plateau phase. The rest-frame zero-velocity is marked with a grey line. The red lines mark the absorption-minima of the P-Cygni absorption (8,500 $\rm km\ s^{-1}$) and the high-velocity absorption (13,500 $\rm km\ s^{-1}$) when they first appear simultaneously in the spectra on $\sim$\,16\,d. The blue edge of the high-velocity feature (20,000 $\rm km\ s^{-1}$) is shown with a blue line. The clumpy features that appear over several epochs are indicated in green boxes.}
    \label{fig:halpha}
\end{figure*}

The complete photospheric phase spectroscopic evolution of SN~2023ixf from 7\,d to 82\,d is presented in Figure~\ref{fig:specp}. During the transition from the flash phase to the photospheric phase, we see the reminiscence of interaction with the dense CSM in the form of intermediate-width Lorentzian emission along with the emergence of broad P-Cygni features of $\rm H\alpha$ and $\rm H\beta$ \citep{2023teja}. This spans until $\sim$\,10\,d, beyond which the Lorentzian emission profile vanishes. 

Beginning $\sim$\,16\,d, we observe a more intricate and complex multi-peaked profile of $\rm H\alpha$ with a persistent high-velocity absorption feature, which we extensively discuss in Section~\ref{sec:halpha}. In addition to the distinguishable Balmer lines in the early spectra, several distinct features start to appear $\sim$\,16\,d onwards, particularly \ion{Ca}{2} H\&K and \ion{He}{1} 5876 \AA. As the photosphere recedes further into the ejecta, we detect the appearance of various lines of \ion{Fe}{2}, \ion{Sc}{2} and \ion{Ba}{2}, and the \ion{Ca}{2} Triplet beginning $\sim$\,25\,d. These metal features are visibly developed around $\sim$\,36\,d. The broad \ion{Na}{1D} starts to emerge at the location of \ion{He}{1} $\lambda$ 5876, evident from the slightly broadened absorption trough at around $\sim$\,40\,d (continuing up to $\sim$\,54\,d). This is further indicative of the cooling of the ejecta \citep{2017gutierrez} and is suggestive of an ejecta that is cooling a lot slower than typical SNe II as the appearance of \ion{Na}{1D}, is usually observed around $\sim$\,30\,d \citep{2017gutierrez}. Towards the end of the plateau drop, i.e. around \,70\,d, we also begin to observe \ion{O}{1}~$\lambda$~7774. 

In the top-right and bottom-right panels of Figure~\ref{fig:specp}, we compare the spectrum of SN~2023ixf with other SNe II spectra encompassing a sample of SNe IIL and short plateau SNe i.e. SN~2013by \citep{2015valenti}; SN~2014G \citep{2016terreran}; SN~2018gj \citep{2023teja18gj}; SN~2020jfo \citep{2022teja}; SN~2020pni \citep{2022terreran}; SN~2020tlf \citep{2022galan}. Photospheric features develop pretty late in SN~2023ixf compared to other Type IIL and short-plateau SNe II, significantly later than normal SNe II. In the early plateau phase ($\sim$\,25\,d), the spectra blueward of H$\alpha$ is generally dominated by metallic features as in the case for SN~2014G, SN~2020jfo and SN~2020pni; however, the features are relatively underdeveloped in SN~2023ixf and SN~2018gj. The line strengths of absorption features appear much weaker in SN~2023ixf than in spectral lines of other SNe. This plausibly hints at the ejecta being still hot and/or metal-poor compared to normal SNe II. We also observe that SN~2023ixf exhibits a weaker $\rm H\alpha$ P-Cygni absorption during its early photospheric phase in tandem with short-plateau SNe from \citep{2021Hiramatsu} but not with short-plateau SNe 2018gj and 2020jfo. This observation suggests that weaker Balmer absorption doesn't necessarily result from a lower mass of the hydrogen envelope in short-plateau SNe. The weaker Balmer absorption is also seen in SNe 2014G and 2020pni in our comparison sample. The spectral features are likely less prominent due to the luminosity from interaction enhancing the ejecta's temperature (and ionization), leading to an ionization wave penetrating the ejecta inwards from the cold, dense shell (CDS, \citealp{1994chevalier}). CDS arises from the higher density of the inner CSM ($\rm >\ 10^{-14}\ g\ cm^{-3}$, see Section~\ref{sec:modelmesa}) resulting from the cooling of shocked areas during the early phases. This phenomenon aligns with observations of Type IIL SNe, which typically exhibit weaker P-Cygni absorption troughs since these SNe likely have more significant interaction with CSM than typical Type IIP SNe \citep{2014gutierrez, 2016polshaw}.


\subsection{Evolution of \texorpdfstring{H$\rm\alpha$}{Halpha} : High-Velocity Absorption and Clumpy CSM}
\label{sec:halpha}


We show the temporal evolution of the $\rm H\beta$ and $\rm H\alpha$ covering the photospheric phase of SN~2023ixf in Figure~\ref{fig:halpha}. We observe the emergence of the broad P-Cygni feature of $\rm H\alpha$ by 7\,d in our spectral sequence with a blue edge of $\sim$\,8,500 $\rm km\ s^{-1}$. A similar broad P Cygni absorption manifests around $\rm H\beta$, confirming the emergence of the SN ejecta. A few epochs later, we observe a high-velocity (HV) broad absorption trough the blue-ward of H\,$\rm \alpha$ P-Cygni absorption beginning $\sim$\,16\,d where the two features are merged and cannot be distinguished. This merged broad component slowly migrates to two distinct components in the spectrum on $\sim$\,27 d. The appearance of the HV absorption in H$\rm \alpha$ is synonymous with the \textit{red peak} in light curves of SN~2023ixf. The absorption minima of the HV absorption lies roughly at 13,500 $\rm km\ s^{-1}$ whereas the blue edge of the absorption profile extends up to 20,000 $\rm km s^{-1}$ at $\sim$\,16\,d. We confirm its association with hydrogen since we see an analogous profile in $\rm H\beta$, although its effect is not as pronounced due to the low optical depth of these lines. This HV feature was also reported by \citet{2023teja}, who dismissed its association with \ion{Si}{2} as it would lead to line velocities lower than the photospheric velocity. 

In the case of an expanding SN ejecta, we tend to observe an absorption component forming from the inner layers of the ejecta moving towards our line-of-sight, leading to a P-Cygni profile of H$\rm\alpha$ during the plateau phase. The outer recombined ejecta does not contribute towards the absorption profile \citep{2007chugai}. However, the collision between the SN ejecta and CSM creates a dual-shock structure, where the forward shock moves through the CSM while the reverse shock travels within the SN ejecta \citep{1982chevalier}. This results in the ionization of the outer layers of the unshocked ejecta by Lyman-$\rm \alpha$ photons, causing the emergence of HV absorption features blueward of the P-Cygni absorption \citep{2007chugai}. Such an HV feature of hydrogen arising due to interaction is generally narrow and typically starts appearing a month after the explosion \citep{2017gutierrez}. Such HV features are generally narrow and don't show a considerable evolution in velocity. In contrast, the HV absorption feature in SN~2023ixf is broad (FWHM $\sim >$ 7,000 $\rm km\ s^{-1}$), starts appearing at 16\,d and shows considerable evolution in velocity from $\sim$\,13,500 $\rm km\ s^{-1}$ at 16\,d to $\sim$\,9,500 $\rm km\ s^{-1}$ at 70\,d. Hence, it is unlikely that the feature arose due to the interaction with the dense CSM. We discuss later in Section~\ref{sec:csmgeometry} that the feature likely arises from the freely expanding ejecta across the polar latitudes of the SN, whereas the P-Cygni profile arises from the decelerated ejecta breaking out of the dense CSM.

The P-Cygni profile of H$\rm \alpha$ from $\sim$\,10 -- 32\,d also shows many intricate structures indicating the presence of clumpy matter in the interaction region. We spot many distinct clumpy features at similar velocities in H$\rm \alpha$ and H$\rm \beta$, and they disappear as the SN evolves into the mid-plateau phase ($>$\,40\,d), indicating that the SN ejecta overcomes most of the clumps beyond the mid-plateau phase. 

The emission peak of $\rm H\alpha$ has a blueshifted offset by as much as $\sim$\,3,000 $\rm km\ s^{-1}$ at $\sim$\,32\,d but evolves towards zero rest velocity by the end of the plateau. This is likely a result of the steep density profile of the ejecta leading to a higher occultation of the receding portion of the ejecta \citep{2014banderson}. The blueshifted offset of the emission peak of H$\rm \alpha$ (at 30\,d) correlates with the decline rate during the plateau phase and the peak luminosity of SNe II \citep{2014banderson}. The steep decline rate of SN~2023ixf ($\rm s_2\,\sim\,1.9\,mag\ (100\,d)^{-1}$ is in agreement with its large offset of $\sim$\,3,000 $\rm km\ s^{-1}$ indicating that the ejecta mass is smaller in SN~2023ixf leading to its shorter plateau. Although this effect is commonly seen in SNe II, this effect is more pronounced in only a handful of events, e.g., SN~2014G \citep{2016terreran} and SN~2018gj \citep{2023teja18gj}, where the emission peak stays blue-shifted late into the nebular phase.

\subsection{Line velocity Evolution}
\label{sec:photvel}

\begin{figure}
\centering
	 \resizebox{\hsize}{!}{\includegraphics{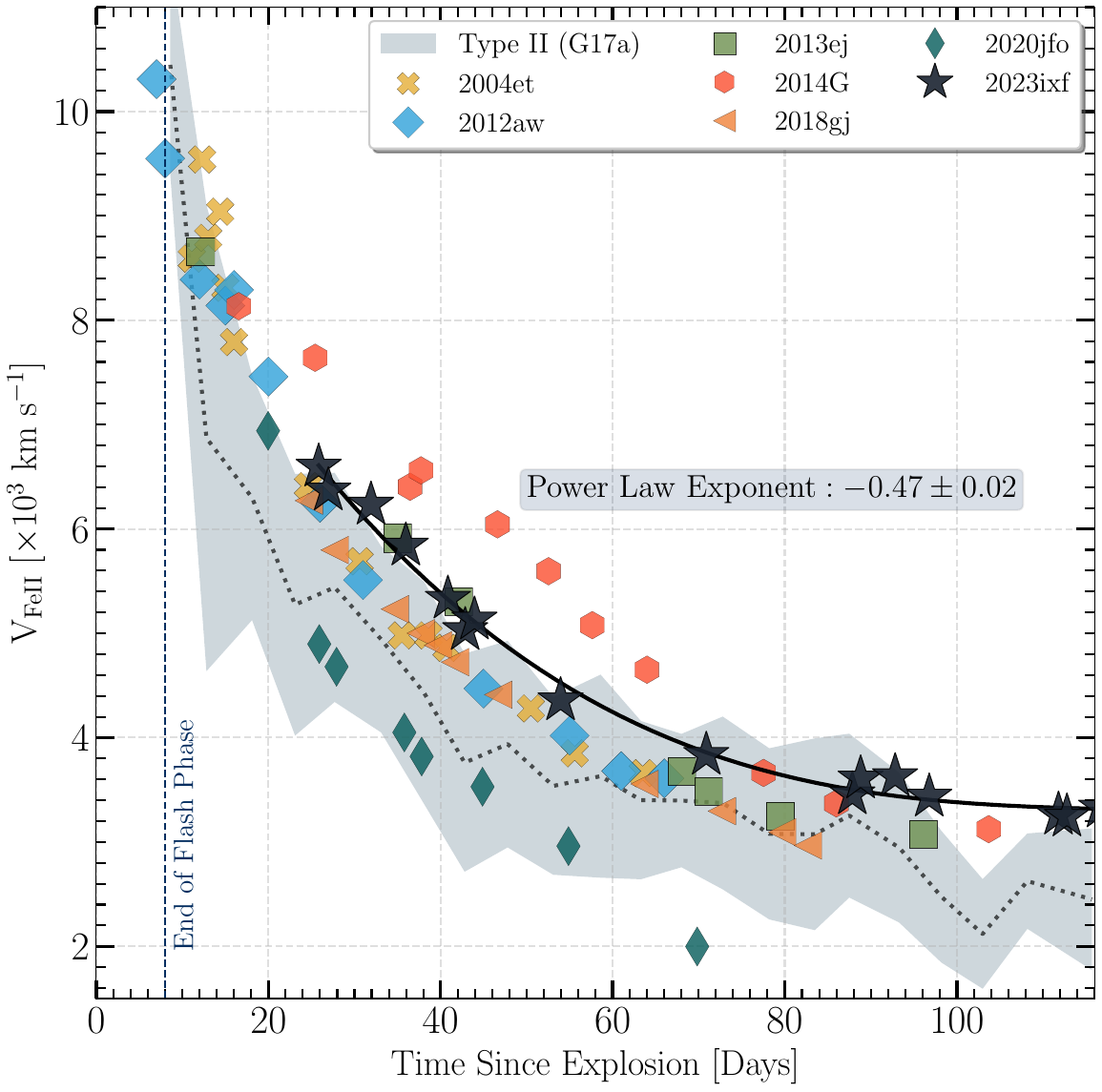}}
    \caption{Comparison of \ion{Fe}{2} $\lambda$ 5169 line velocity evolution of SN~2023ixf in comparison with SNe II from the literature and the mean photospheric velocity evolution for SNe II computed by \citet{2017gutierrez}.}
    \label{fig:compphotvel}
\end{figure}

We compare the velocity evolution of \ion{Fe}{2} $\lambda$ 5169 for SN~2023ixf with other SNe II in Figure~\ref{fig:compphotvel}. We estimated the line velocity evolution from the blue-shifted absorption trough of their line profiles in the redshift-corrected spectral sequence of SN~2023ixf. We adopt the \ion{Fe}{2} $\lambda$ 5169 velocities as the photospheric velocity since it forms closest to the photosphere \citep{2005adessart} and is the least blended among the iron lines seen in our spectral sequence. The photospheric radius estimated from \ion{Fe}{2} absorption trough closely mirrors the blackbody radius as shown in Figure~\ref{fig:bollc}. During the mid-plateau phase ($\sim$\,53\,d), the \ion{Fe}{2} $\lambda$ 5169 velocity inferred for SN~2023ixf is 4350 $\rm km\ s^{-1}$. This positions it at the higher end of the 1-$\sigma$ range when compared with the mean velocity of \ion{Fe}{2} $\lambda$ 5169 for an extensive collection of SNe II by \citet[3537\,$\pm$\,851 km/s]{2017gutierrez}. 
This is evident in the spectral comparison during the late-plateau phase (see Figure~\ref{fig:specp}) since we observe higher line blending in SN~2023ixf due to its higher photospheric velocity in comparison to all other SNe II in our comparison except that of SN~2014G.

The trend continues onto the early nebular phase ($\sim$\,116\,d), where SN~2023ixf displays a \ion{Fe}{2} $\lambda$ 5169 velocity of 3330 $\rm km\ s^{-1}$ compared to the mean value of 2451\,$\pm$\,679 $\rm km\ s^{-1}$ from \citet{2017gutierrez}. The higher photospheric velocity is in agreement with the luminosity-velocity correlation of homologously expanding recombination front of hydrogen \citep{2009kasen} since SN~2023ixf is brighter during the mid-plateau phase compared to a normal Type II SN \citep{2014anderson}. Additionally, the CSM interaction in SN~2023ixf could also drive the elements at outer/faster regions of the ejecta to be reionized and recombined, leading to a higher estimate of photospheric velocity \citep{2019andrews}.

\begin{figure*}[hbt!]
\centering
	 \resizebox{\hsize}{!}{\includegraphics{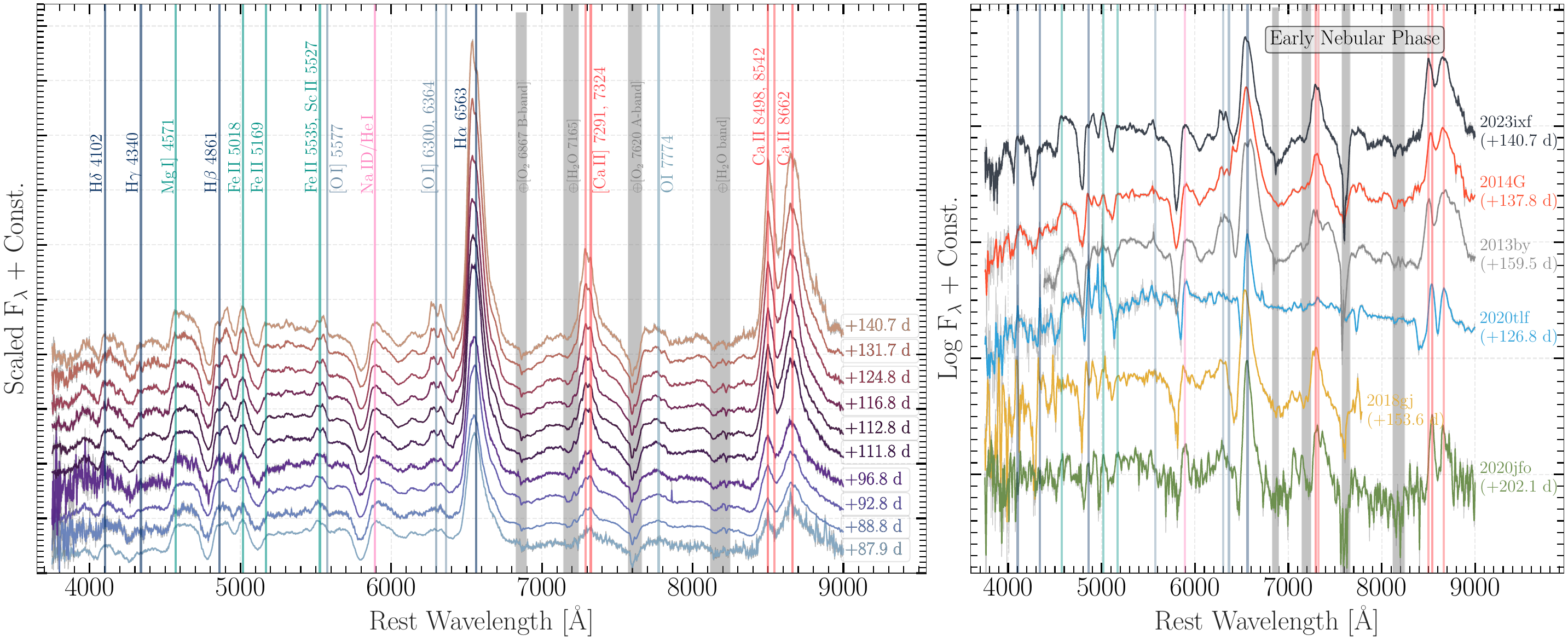}}
    \caption{\texttt{Left Panel}: Nebular spectroscopic evolution of SN~2023ixf from HCT. The marked vertical lines indicate the rest wavelength of the labeled spectral features. \texttt{Right Panel:} Comparison of the early and late plateau phase spectrum of SN~2023ixf with other SNe II from the literature with signatures of CSM interaction in the top and bottom panels, respectively. References: SN~2013by \citep{2015valenti}; SN~2014G \citep{2016terreran}; SN~2018gj \citep{2023teja18gj}; SN~2020jfo \citep{2022teja}; SN~2020tlf \citep{2022galan}. (The spectroscopic data is available as data behind the figure.)}
    \label{fig:specn}
\end{figure*}

\subsection{Early Nebular Phase \texorpdfstring{$>$}\ 90~d}
\label{sec:specneb}

\begin{figure}[hbt!]
\centering
	 \resizebox{\hsize}{!}{\includegraphics{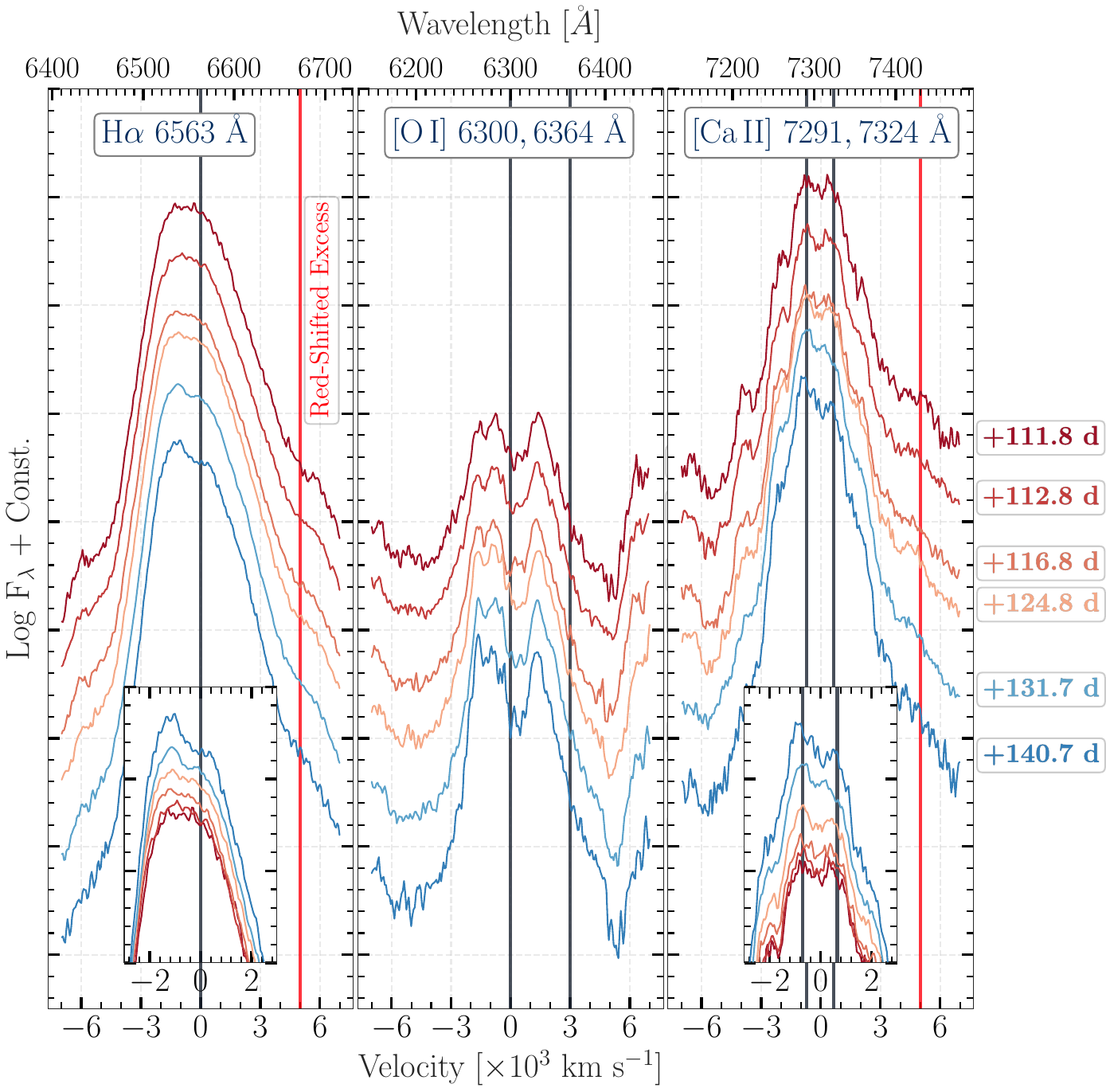}}
    \caption{Evolution of H$\rm \alpha$ 6563 \AA, [\ion{O}{1}] 6300, 6364 \AA, and [\ion{Ca}{2}] 7291, 7324 \AA\ during the early nebular phase of SN~2023ixf. The vertical lines represent the rest-wavelength of the features labeled in each subplot. The zero-velocity corresponds to 6563\AA, 6300 \AA, and 7308\AA\ in the 3 subplots, respectively. The blueshifted offset of 1500 $\rm km \ s^{-1}$in \ion{O}{1} doublet is distinctly visible. The redshifted excess in H\,$\rm \alpha$ and [\ion{Ca}{2}] $\rm \lambda \lambda$7291,\,7324 is labelled at +5,000 $\rm km\ s^{-1}$. The insets show a zoomed view of the peak of the H$\rm \alpha$ and [\ion{Ca}{2}], showing redward attenuation as the SN progressed onto the nebular phase.}
    \label{fig:nebularlines}
\end{figure}

The nebular-phase spectral sequence of SN~2023ixf is shown in the left panel of Figure~\ref{fig:specn}. The nebular spectra of SN~2023ixf displays prominent emission features of H\,$\rm \gamma$, \ion{Mg}{1}] $\lambda$ 4571, H\,$\rm \beta$, \ion{O}{1} $\lambda$ 5577, \ion{Na}{1}\,D / \ion{He}{1}, [\ion{O}{1}] $\rm \lambda \lambda$6300,\,6364, H\,$\rm \alpha$, [\ion{Ca}{2}] $\rm \lambda \lambda$7291,\,7324 and the \ion{Ca}{2} NIR triplet with a flat continuum, typical of SNe II. We observe signs of asymmetry in the line profile of certain emission features during the nebular phase. We observe a dual-peaked axisymmetric\footnote{The separation between the two components is less than that of the two components of the \ion{O}{1} doublet} profile of the [\ion{O}{1}] doublet blueshifted by 1500 $\rm km \ s^{-1}$ in SN~2023ixf during the early nebular phase in Figure~\ref{fig:nebularlines}. In addition, we observe an apparent redshifted excess in H\,$\rm \alpha$ and [\ion{Ca}{2}] $\rm \lambda \lambda$7291,\,7324 at around +5,000 $\rm km\ s^{-1}$ possibly indicating asymmetries in the ejecta. However, the H$\rm \alpha$ and [\ion{Ca}{2}] showed a single peak symmetric profile with visible signs of redward attenuation as the SN progressed into the nebular phase. 

The emergence of an asymmetric emission profile of H$\rm\alpha$ and [\ion{Ca}{2}] arising due to the attenuation of the red-ward emission from the receding portions of the ejecta is first noticed in our spectral sequence beginning $\sim$\,125\,d in Figure~\ref{fig:nebularlines}. This indicates the onset of dust formation and was first noticed in SN~1987A \citep{1989lucy}. We also observe that the red-blue asymmetry increases as the SN evolves into the nebular phase, indicating an increased dust formation with time \citep{2019bevan}. The early signatures of dust suggest its formation is happening inside the CDS since the SN ejecta during the 125 -- 140\,d is too warm for the condensation for molecules \citep{1991kozasa}. The flash-ionization features in the early spectral sequence and the steep rise in early UV light curves of SN~2023ixf conclusively indicated the presence of a dense CSM. As the shockwave from the SN encountered the denser CSM, it decelerated, compressing the material and increasing the density within the shocked CSM. Radiative cooling then facilitated the emission of photons, aiding in its cooling and forming a distinctive CDS \citep{2009chugai} in the denser CSM. The CDS enables an additional pathway for dust formation in interacting SNe II \citep{2018rho}. In addition, the clumpiness within the extended CSM encompassing SN~2023ixf facilitates the formation of additional CDS (in addition to its formation in the dense CSM), consequently enhancing molecule formation and eventually forming dust \citep{2011inserra}. Since nebular phase H$\rm \alpha$ arises from the inner ejecta, it wouldn't show wavelength-dependent attenuation if the dust is formed in the outer CDS. This suggests that the regions where Balmer lines form and dust formation occurs essentially overlap, indicating thorough mixing of the CDS into the inner ejecta following a significant episode of CSM interaction \citep{2019bevan}. 

We also see evidence of flattening in the $Ks$-band light curve of SN~2023ixf beyond 125\,d in Figure~\ref{fig:applc} evolving at 1.3\,$\pm$\,0.1 $\rm mag\ {100\,d}^{-1}$ against the relatively consistent decline of 1.8\,$\pm$\,0.1 $\rm mag\ {100\,d}^{-1}$ in the $J$ and $H$ bands. This indicates that the continuum luminosity in NIR is evolving steadily; however, the $Ks$-band light curve is evolving rather slowly due to the emission from CO overtone around 2.3$\mu$m and/or formation of warm dust. Although we don't have spectral conformation in NIR, Type II SN~2017eaw \citep{2018rho} showed the presence of the first overtone of CO as early as 124\,d and also showed flattening in their $Ks$-band light curves. This strengthens our inference for indirect conformation of molecular CO formation and eventually dust in the case of SN~2023ixf. 

\section{Polarimetric Analysis}
\label{sec:pol}

\begin{figure*}
\centering
	 \resizebox{0.49\hsize}{!}{\includegraphics{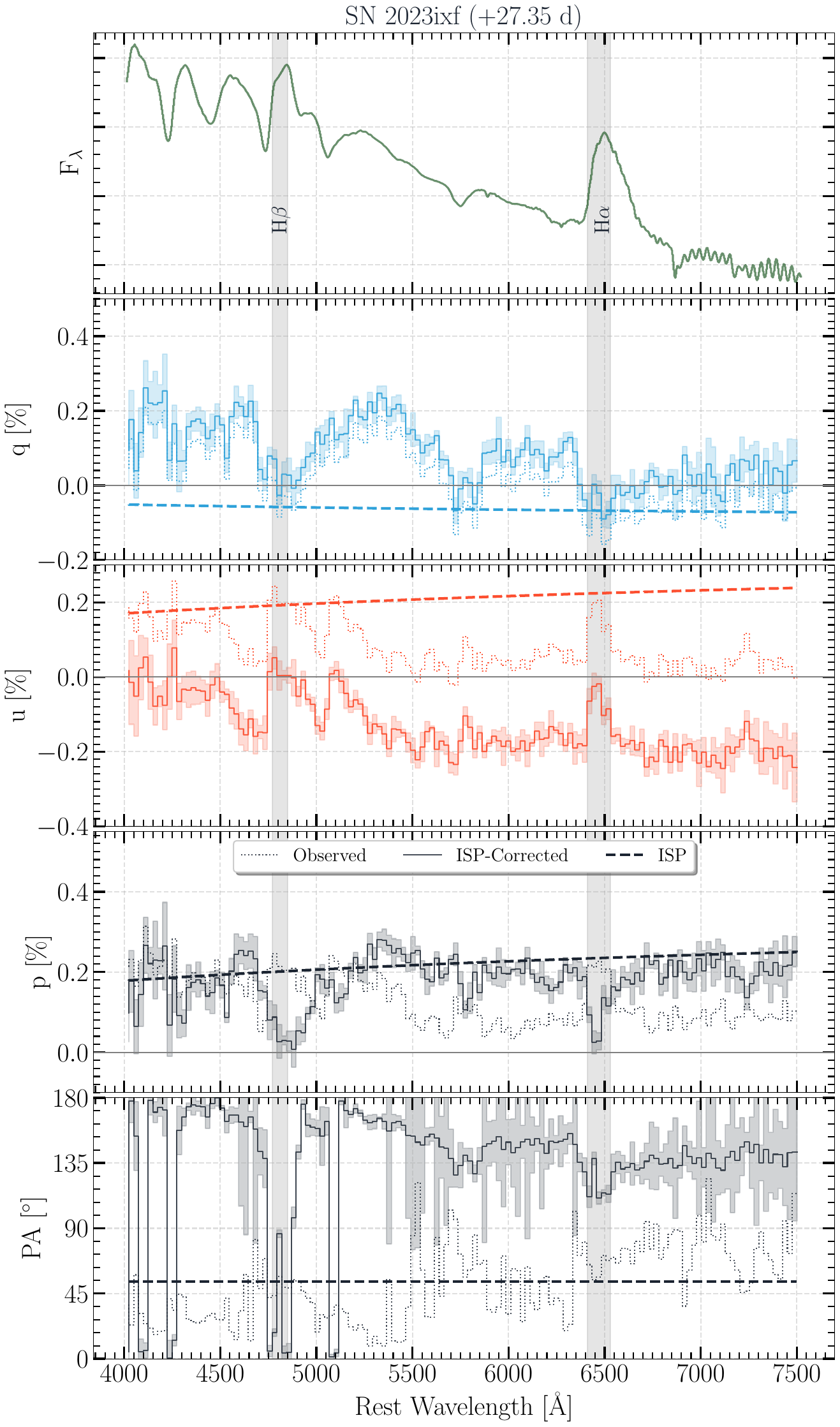}}
  \resizebox{0.49\hsize}{!}{\includegraphics{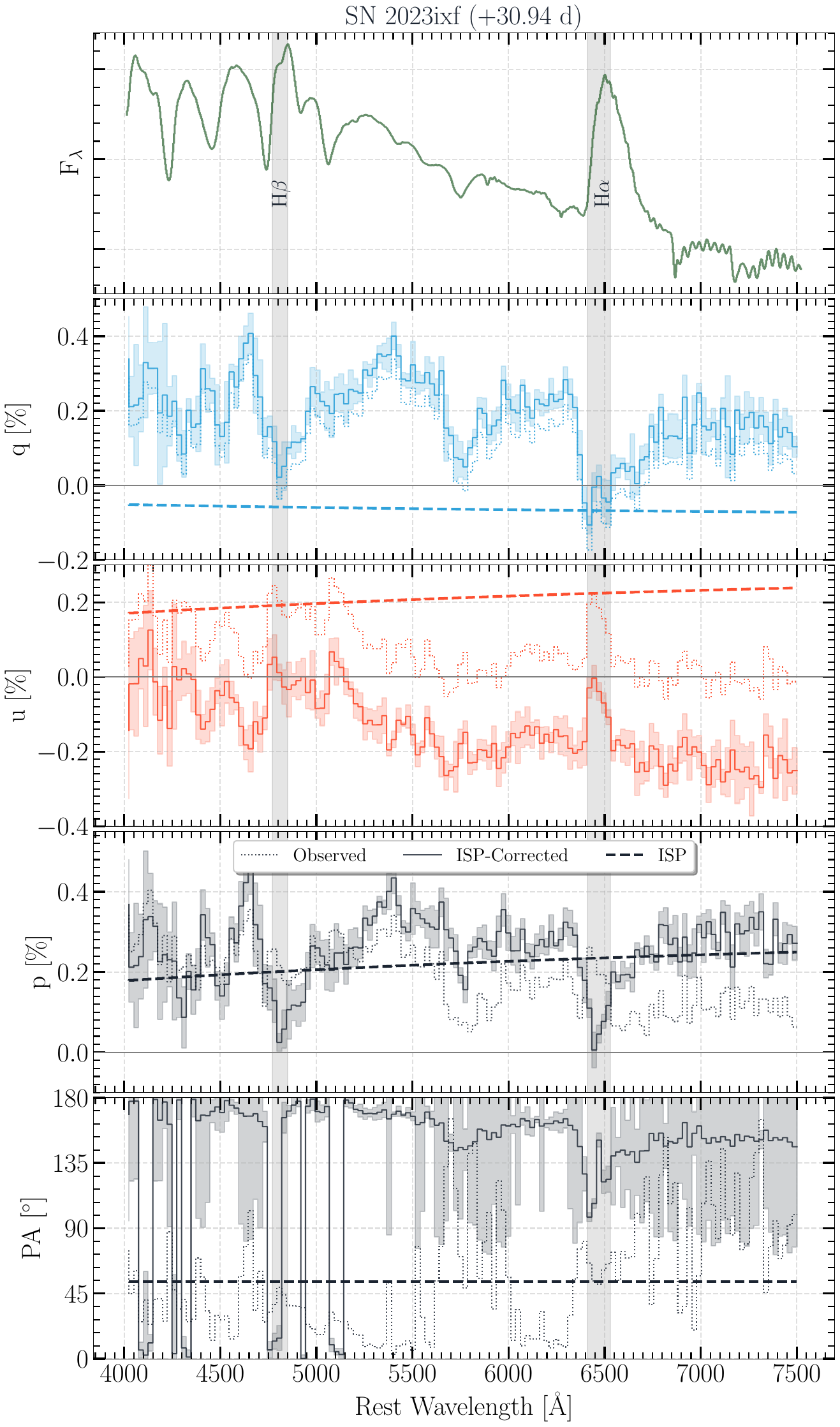}}
    \caption{Spectropolarimetric observation of SN~2023ixf averaged out over a) 27.9\,d (June 14,15,16) and b) 30.9\,d (June 18, 19). The five panels from the top represent the flux spectrum ($\rm F_\lambda$), the fractional Stokes parameters \textit{q} and \textit{u}, polarization (\textit{p}), and PA, respectively and binned to 25 \AA\ for clarity. The panels show the observed and the ISP-corrected polarization along with the estimated ISP. The prominent emission features of H$\rm\alpha$ and H$\rm\beta$, chosen for ISP estimation, are indicated by the vertical shaded region. The polarization uncertainty is shown via a shaded region around the solid line. (The continuum polarization data is available as data behind the figure.)}
    \label{fig:specpol}
\end{figure*}

Figure~\ref{fig:specpol} presents spectropolarimetry observations of SN~2023ixf during the early plateau phase. Due to the low signal-to-noise ratio of the individual epochs, we averaged out the first three epochs (June 14, 15, 16) and the following two epochs (June 18, 19) into two spectropolarimetric observations represented by their average phase of observations. The polarization is expressed as a sum of the $q$ and $u$ stokes parameters, P\,$=$\,$\sqrt{q^2 + u^2}$, and the polarization angle (PA) is given by the PA\,=\,0.5 \,$\tan^{-1} \left(u/q \right)$. 

\subsection{Estimating Interstellar Polarization}
\label{sec:isp}

To analyze the intrinsic polarization of SN~2023ixf, it's essential to quantify the Interstellar Polarization (ISP) originating from the irregularly shaped dust particles present across the line of sight in both the Milky Way and the host galaxy M~101. To estimate an upper limit on the ISP, we assume a Serkowski-Galactic type ISP \citep{1975serkowski} for M~101 by assuming a similar size and dust composition as the Milky Way. Using the extinction values along the line-of-sight of SN~2023ixf in Section~\ref{sec:host}, the upper limit on the ISP from \citet{1975serkowski} is 9 $\times$ $E(B-V)$ = $\rm P^{ISP}_{Max}$ $\sim$ 0.35\%. Since the three peaks in our imaging polarization evolution of SN~2023ixf are all higher than the maximum limit on ISP, we reasonably conclude that the observed polarization characteristics originate from the SN itself.

However, to infer any further detail from the polarimetric and spectropolarimetric data, especially during the mid-plateau phase, where the mean continuum polarization seems to drop below the $\rm P^{ISP}_{Max}$, we must constrain ISP more rigorously from our observed dataset. To extract ISP from our dataset, we assumed that emission peaks of P-Cygni profiles have no intrinsic polarization, similar to several spectropolarimetric studies \citep{2006chornock, 2019nagao}. In the case of optically thick lines such as H$\rm \alpha$, multiple scattering processes diminish the geometric characteristics conveyed by the photospheric radiation from the continuum, consequently leading to the depolarization of the emission line \citep{1996hoflich}. The emission peak of H$\rm \alpha$ and H$\rm \beta$ in Figure~\ref{fig:specpol} exhibits non-zero polarization, featuring an angle distinct from that measured in the continuum range. We estimated the wavelength dependence of the ISP using the averaged spectropolarimetric observations of our entire data set (June 14, 15, 16, 18 and 19) by fitting the wavelength ranges around the peak of H$\rm \alpha$ and H$\rm \beta$ with the Serkowski function \citep{1975serkowski} i.e. $\rm P(\lambda) = P_{max} \exp (-K \ln 2 (\lambda_{max}/ \lambda))$. We fixed the K value to 1.15 (same as Milky Way) and obtained a best-fit for the ISP at $P_{max}\,\sim\, 0.37 \%$ and $\lambda_{max} \,\sim\, 3650 \, \text{\AA}$. At the rest-wavelength of 6000\AA\, we estimate an ISP of $\rm q_{ISP}$\,=\,-0.07\,$\pm$\,0.02 \%, $\rm u_{ISP}$\,=\,0.22\,$\pm$\,0.06 \% and use this to correct the ISP in our $R$-band imaging polarization evolution and values adopted from \citet{2023vasylev} in Figure~\ref{fig:pollc}.

\subsection{Polarization in the Stokes Q-U Plane and Comparison with other SNe II}
\label{sec:comppol}

\begin{figure}
\centering
	 \resizebox{\hsize}{!}{\includegraphics{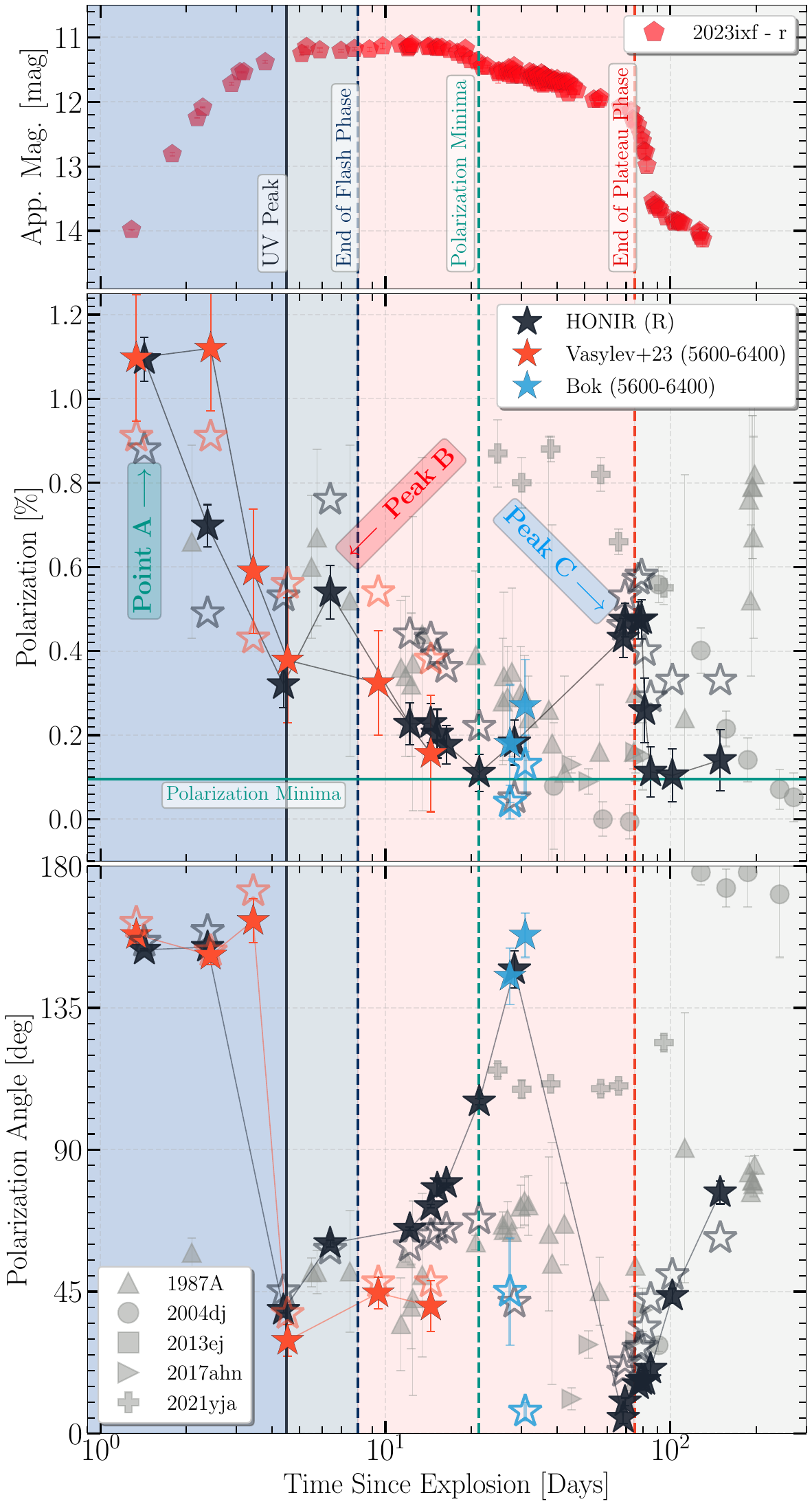}}
    \caption{Temporal evolution of polarization and its PA in $R$-band for SN~2023ixf in comparison to SN~1987A \citep{1988barrett}, SN~2004dj \citep{2006leonard} and SN~2013ej \citep{2016brajesh}. Estimates of continuum polarization from \citet{2023vasylev} for SN~2023ixf are overplotted. The observed and ISP-corrected polarization is shown with empty and filled markers, respectively. The $R$-band light curve of SN~2023ixf is plotted in the top panel for reference.}
    \label{fig:pollc}
\end{figure}

\begin{figure}
\centering
	 \resizebox{\hsize}{!}{\includegraphics{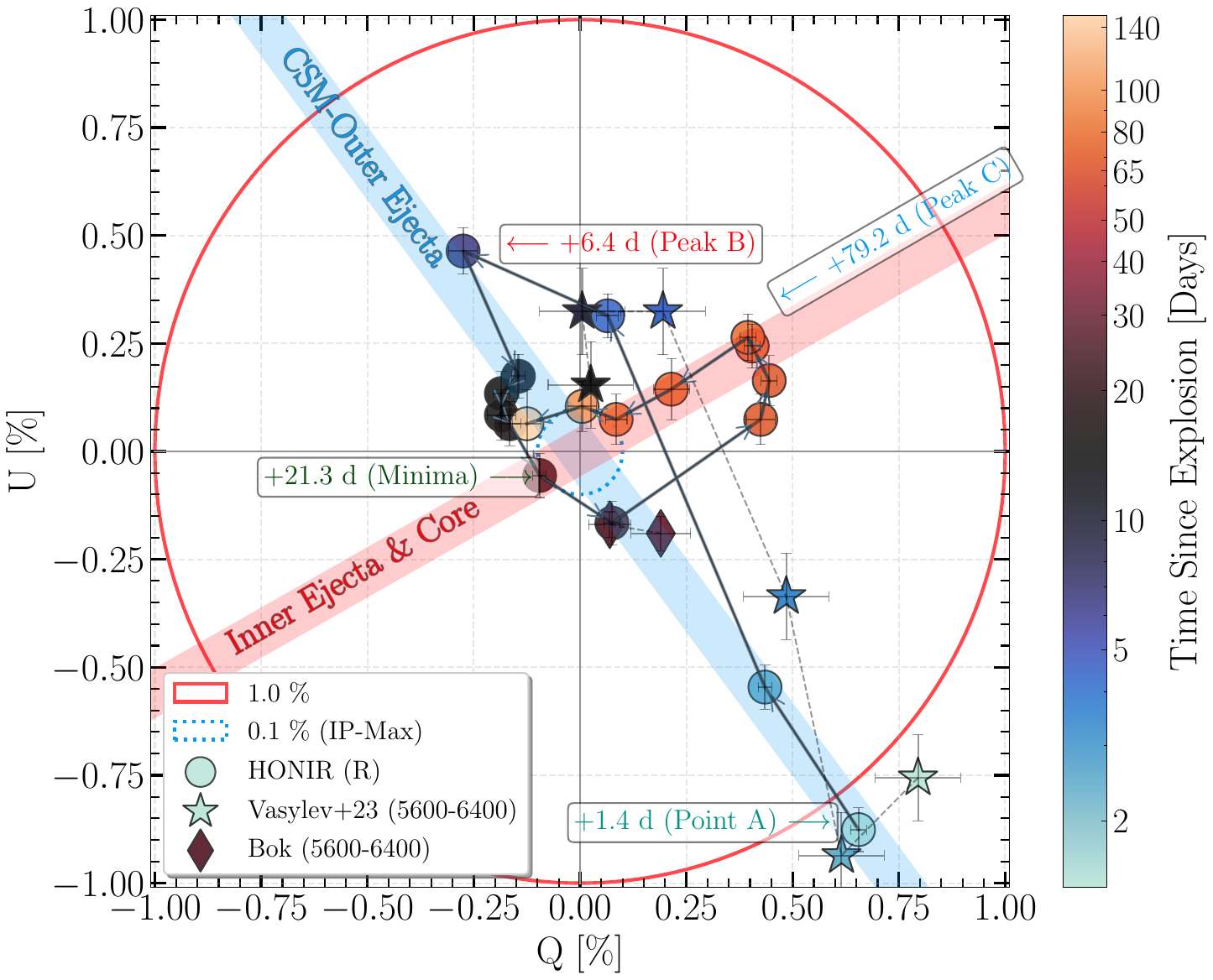}}
    \caption{The observed Stokes parameters from HONIR and Bok in the Q-U plane for SN~2023ixf. The color map on the right represents the SN phase. The stokes Q-U from \citet{2023vasylev} for SN~2023ixf is overplotted. The shaded regions depict the dominant axis of the CSM-outer ejecta (in blue) and the inner ejecta-core (in red) inferred from the temporal evolution of SN~2023ixf.}
    \label{fig:qupol}
\end{figure}

In Figure~\ref{fig:pollc}, we present ISP-corrected imaging polarization evolution of SN~2023ixf in $R$-band spanning $\sim$\,1.4\,d until 140\,d. Figure~\ref{fig:pollc} shows 3 distinct peaks in polarization; Point A\footnote{We refer to the first observation as point A since we don't have any data prior to this epoch.} [$(1.09 \pm 0.05)\%$, $(153.4\pm 0.3)^{\circ}$] at 1.4\,d, Peak B [$(0.54 \pm 0.06)\%$ at 6.4\,d, $(60.3\pm 1.1)^{\circ}$] and Peak C [$(0.48 \pm 0.05)\%$, $(16.9\pm 0.8)^{\circ}$] at 79.2\,d. The observed stokes parameters ($Q_{obs}$ and $U_{obs}$) in $R$-band are shown in Figure~\ref{fig:qupol}. Our first imaging polarimetric observation of SN~2023ixf was conducted at 1.4\,d, corresponding to Point A, and is associated with the flash-ionized phase due to the interaction of the SN ejecta with an asymmetric, compact, and dense CSM. This is similar to the value of continuum polarization inferred using spectropolarimetric data by \citet{2023vasylev}. However, we observe a decrease in the polarization signature observed from the dense CSM on the 2nd epoch, i.e., $\sim$\,2.5\,d, which is in stark contrast to the rising luminosity (see a six-fold rise in $R$-band luminosity). However, the continuum polarization reported by \citet{2023vasylev} at the same epoch, after correction for ISP, indicates a slight polarisation increase from the first epoch, consistent with rising luminosity driven by interaction. The discrepancy could likely be due to the increased line depolarization of H$\rm \alpha$, leading to a decrease in $R$-band polarization as the PA remains the same. 

At the next epoch on $\sim$\,4.4\,d, we see a change in PA to $\sim$\,55\,$^{^\circ}$ which later evolves to the polarization peak B at 6.4\,d. The polarization peak B lies towards the end of the flash phase transitioning to the early photospheric phase, which shows interaction signatures in the form of intermediate-width P-Cygni profiles \citep{2023smith} and close to the emergence of the broad Balmer features at 7\,d (see Section~\ref{sec:specphot}). Polarization at Peak B likely arises from interaction with extended CSM and/or an aspherical shock structure of SN~2023ixf. Following Peak B, we see a decline in polarization across the early plateau phase ($<$\,22\,d); however, the PA continues evolving slowly. There is slight disagreement with the continuum PA reported by \citet{2023vasylev} likely due to the depolarization of the emission peak of the P-Cygni profile of H$\rm \alpha$. We do observe a revival in the polarization evolution at 28\,d with the PA evolving from $\sim$\,100\,$^{^\circ}$ to $\sim$\,150\,$^{^\circ}$. We have a 40-day gap before our next observations at 70\,d, before the end of the plateau phase. The observed polarization evolution of SN~2023ixf after 70\,d demonstrated a rise towards Peak C in polarization at 79.2\,d towards the end of the plateau phase. Since the innermost layers of the ejecta ($<$ 3000 $\rm km\ s^{-1}$) are revealed as a result of the entire hydrogen envelope becoming optically thin, this suggests that the electron-scattering atmosphere of the inner core shows a substantial departure from spherical symmetry. The polarization dropped slowly across the early nebular phase, whereas the PA increased gradually. SN~2023ixf shows two preferred dominant axes: a) along the CSM-outer ejecta, where the SN evolved during the early phase until the epoch of polarization minimum on 21.3\,d, and b) along the inner ejecta/core, where the SN evolved during the late-plateau and early nebular phase, which is inferred from its polarization evolution.


In a recent study, \citet{2024Nagao} compared the intrinsic polarization properties of 15 SNe II. Broadly, it was inferred that the degree of polarization rises significantly ($\sim\,1.5\%$) in the transition/nebular phase; however, it is usually lower during the photospheric phase ($\sim\,0.1\%$). Anomalous cases exist in SN~1987A \citep{1989jeffery} and SN~2021yja \citep{2024vasylev}, where a significant polarization is observed during the early plateau phase, gradually declining towards the end of the plateau phase. This is likely driven by electron scattering from interaction with an aspherical CSM, which is analogous to the early continuum polarization (Point A) in SN~2023ixf. However, we also see a secondary peak (Peak B) in polarization in SN~2023ixf, which could possibly arise from interaction with a clumpy low-density extended CSM and/or an aspherical shock structure (see Section~\ref{sec:csmgeometry} for further discussions). Towards the end of the plateau phase, the swift increase in polarization during the late-plateau phase and steady decrease during the early nebular phase is similar to that seen in several SNe II like SN~1987A \citep{1989jeffery}, SN~2004dj \citep{2006leonard} and SN~2013ej \citep{2016brajesh}. Such a polarization evolution is often interpreted as arising from the asymmetries in the inner He-core, and the trend seen is similar in SN~2023ixf. Type IIL SN~2017ahn, which showed early flash-ionization features similar to SN~2023ixf, showed low continuum polarization during the late-plateau phase photospheric and nebular phases, indicating that it might have been asymmetric; however, its viewing angle was lying across its polar axis \citep{2021nagao}. However, it is clear from the observations of SN~2023ixf that the inclination differs from the axis of symmetry.


\section{Light Curve modeling}
\label{sec:modelmesa}

\begin{figure*}
\centering
  	\resizebox{\hsize}{!}{\includegraphics{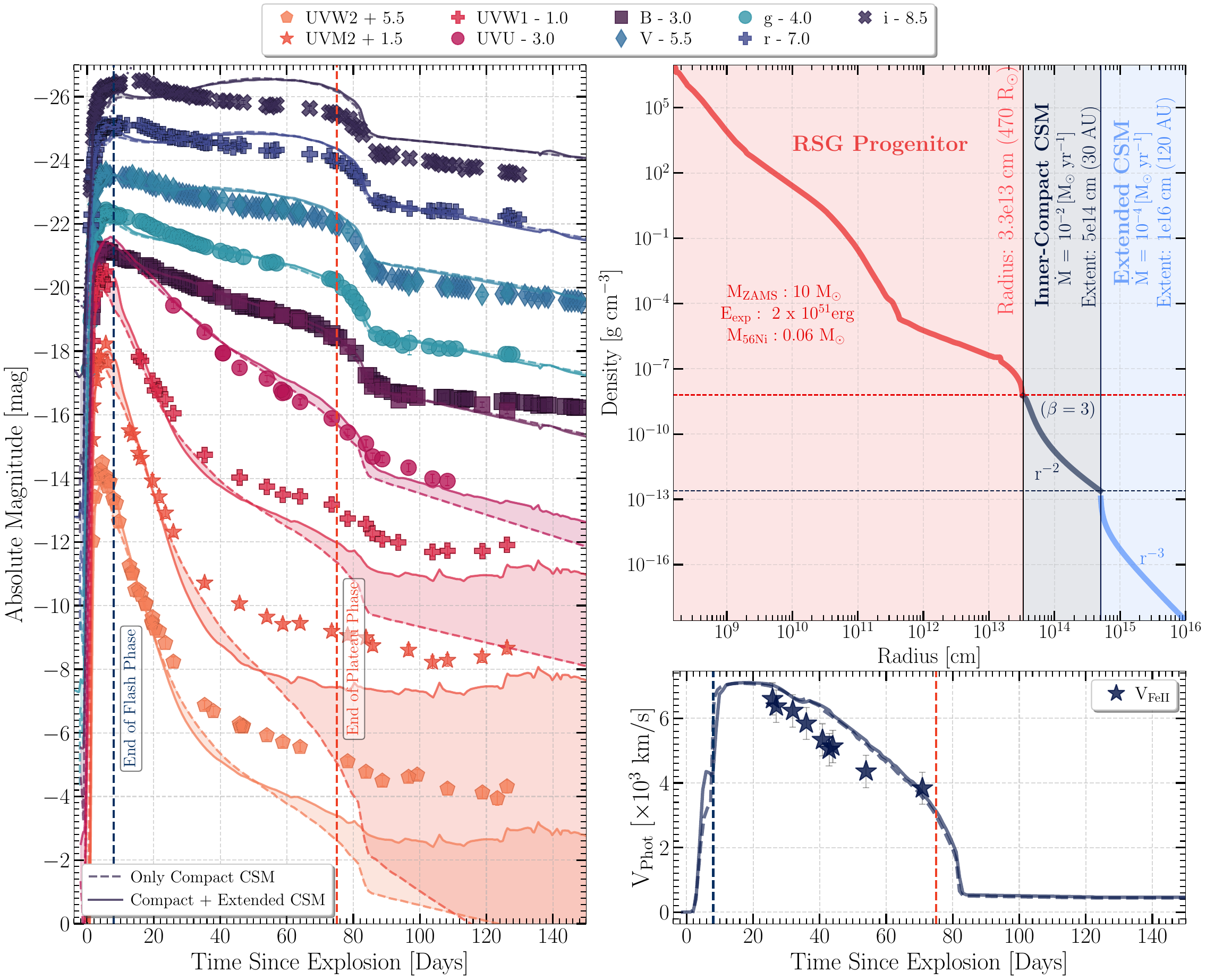}}
    \caption{\textit{Left Panel}: Hydrodynamical modeling of the light curves of SN~2023ixf with only a compact CSM (dashed line) and including an extended CSM (solid line). The shaded region shows the UV excess from the interaction due to the extended CSM. \textit{Top-right panel}: The density structure of the RSG progenitor, the inner-compact CSM, and the extended CSM. The inner-compact CSM extended to $5\times 10^{14}~\mathrm{cm}$ and the extended CSM extends to $1\times 10^{16}~\mathrm{cm}$. \textit{Bottom-right panel}: The photospheric velocity estimates from both the hydrodynamic models are compared with the observed evolution of photospheric velocity.}
    \label{fig:mesamodel}
\end{figure*}

To estimate the explosion parameters and infer the properties of CSM around SN~2023ixf, we performed numerical light curve modeling of SN~2023ixf. We used one-dimensional multi-frequency radiation hydrodynamics code \texttt{STELLA} \citep{1998blinnikov,2000blinnikov,2006blinnikov} for this purpose. \citet{2023moriya} computed light-curve models of SNe II with various progenitor masses, explosion energies, and CSM density profile and its extent using \texttt{STELLA}. The model grid adopts the solar-metallicity RSG progenitor models with the ZAMS masses of 10, 12, 14, 16, 18 $\rm M_\odot$ from \citet{2016sukhbold} and covers explosion energies from $5\times 10^{50}~\mathrm{erg}$ to $5\times 10^{51}~\mathrm{erg}$. The CSM is constructed by assuming the mass-loss rates of $10^{-5}-10^{-1}~M_\odot~\mathrm{yr^{-1}}$ with a terminal wind velocity of $10~\mathrm{km~s^{-1}}$. The density structure of the wind above the progenitors is characterized by the $\beta$ parameter ranging from $0.5$ to $5$. $\beta$ shapes the velocity/density profile of the wind and is inversely proportional to wind acceleration \citep{2017moriya}. We refer the reader to \citet{2023moriya} for additional details regarding the model grid. The extensive details regarding the model comparison will be presented in \citet{2024moriya}.

Using the pre-computed model grid, we first looked for the best matching models in the \textit{g} and \textit{r} bands by searching for models with the minimum $\chi^2$. We found that the models from a low-mass progenitor ($\rm 10~M_\odot$) best fit the light curves of SN~2023ixf. The best fitting models have the explosion energy of around $2\times 10^{51}~\mathrm{erg}$, the $^{56}\,\mathrm{Ni}$ mass of around $\rm 0.06~M_\odot$, the mass-loss rate of around 10$^{-2}$ $\rm M_{\odot}\ yr^{-1}$, the CSM radius of around $6\times 10^{14}~\mathrm{cm}$, and $\beta$ of around $3$. We performed additional numerical light curve calculations to fit the early UV luminosity and found that the model with the CSM radius of $5\times 10^{14}~\mathrm{cm}$ matches the observed dataset, including the photospheric velocity (shown in dashed lines in Fig.~\ref{fig:mesamodel}). The mass of the confined CSM below $5\times 10^{14}~\mathrm{cm}$ in this model is $\rm 0.67~M_\odot$. This extra model assumes that $^{56}\,\mathrm{Ni}$ is uniformly mixed in the entire ejecta. We found that the fully mixed model better fits the transition from the plateau phase to the tail phase. However, they also lead to an excess emission in the redder bands during the late plateau phase as $^{56}\,\mathrm{Ni}$ starts diffusing out earlier than in the case of a centrally concentrated $\rm^{56}$Ni. A much more prominent effect of $^{56}\,\mathrm{Ni}$-mixing was seen in the case of SN~2009ib \citep{2015takats} and SN~2016gfy \citep{2019aavinash}. Since the $\rm^{56}$Ni synthesized in SN~2023ixf is high in comparison to normal SNe II (0.03\,$M_{\odot}$, \citealp{2014anderson}), the late-plateau luminosity bump is much more prominent in the redder bands of our models.

We observed a notable decline in the UV flux around 40 days in the best-fitting model (having only an inner dense CSM) presented above in Fig.\ref{fig:mesamodel}. To address this decrease and account for the sustained UV excess in the late phase of SN~2023ixf, we modeled the light curves with an extended CSM component situated right outside the inner CSM from $5 \times 10^{14}\,\mathrm{cm}$ to $10^{16}\,\mathrm{cm}$. This additional CSM component possesses a mass of 0.025 $\rm M_\odot$ and has a density structure proportional to $r^{-3}$. The low-density extended CSM component helps sustain a prolonged UV brightness until the early nebular phase. Assuming a constant wind velocity of $10~\mathrm{km~s^{-1}}$, the outer CSM component is shaped by an average mass-loss rate of 10$^{-4}$ $\rm M_{\odot}\ yr^{-1}$. The mass-loss rate should increase towards the time of the explosion to form a density structure proportional to $r^{-3}$. We found that the shallower CSM structure, such as that proportional to $r^{-2}$, gives a UV flux excess much higher than observed. Using the observed wind velocities of 50 -- 115 $\rm km\ s^{-1}$ \citep{2023galan, 2023smith, 2023zhang} instead of an assumed value, we infer a mass-loss rate of $0.5 - 1 \times 10^{-1}\ \rm M_{\odot}\ yr^{-1}$ for the inner dense CSM and $0.5 - 1 \times 10^{-3}\ \rm M_{\odot}\ yr^{-1}$ for the extended CSM.


SN~2023ixf has a higher peak luminosity, higher $\rm^{56}$Ni-mass, and a higher photospheric velocity than a typical Type II SN. The peak $V$-band luminosity of SN~2023ixf is $\sim$\,-18.2 mag, which is 4 times brighter than compared to the luminosity of a typical Type II SN, i.e., $\sim$\,-16.7 mag \citep{2014anderson}. The $\rm^{56}$Ni-mass of SN~2023ixf is 0.059\,$\pm$\,0.001 $\rm M_{\odot}$ which is 80\% higher than the $\rm^{56}$Ni-mass estimate of a typical Type II SN, i.e., 0.033 $\rm M_{\odot}$ from \citep{2019anderson}. Previous studies on SNe II \citep{2003hamuy, 2015apejcha, 2017bgutierrez} have shown that more energetic explosions lead to higher photospheric velocities and a higher $\rm^{56}$Ni-yield in the SN. The estimated $\rm^{56}$Ni-mass does indicate that SN~2023ixf is a highly energetic event and is in concordance with the high explosion energy of 2\,$\times$\,10$\rm ^{51}\ erg$ estimated from the light curve modeling. Other works on numerical modeling of SN~2023ixf also suggest high explosion energies. \citet{2024bersten} estimate 1.2 foe as explosion energy for a 12~$\rm M_\odot$ RSG with 10.9~$\rm M_\odot$ as the final progenitor mass. \citet{2023hiramatsu} estimated 1 foe as explosion energy while exploring 2 foe as well for a 12~$\rm M_\odot$ RSG with 11~$\rm M_\odot$ as the final progenitor mass. 


In our case, where we have the best match with a lower mass progenitor, slightly higher explosion energies must be required to match the observed light curves. Although we fit both photospheric velocities and light curves simultaneously, the mass and radius of the progenitors are fixed since we adopt progenitor models from \citet{2016sukhbold}. Given the inherent degeneracies between the ejecta mass, radius, and explosion energy \citep{2019ApJ...879....3GGOLD}, our inferred properties likely represent one possible set of solutions rather than a unique one. However, the high-explosion energy cannot entirely explain the peak luminosity of SN~2023ixf. It is enhanced further by early interaction with the confined CSM. The photospheric velocity of SN~2023ixf is 20\% faster than a proto-typical Type II SN at 50\,d (see Section~\ref{sec:photvel}), and a power-law fit to photospheric velocity evolution returned exponent -0.47\,$\pm$\,0.04, which is slower than the average value derived from a large sample of SNe IIP (--0.581\,$\pm$\,0.034) in \citet{2014afaran}. However, since SN~2023ixf is a short-plateau SN, the slightly low-ejecta mass could be one reason for its higher photospheric velocity \citep{2022teja}.


\section{Discussion}
\label{sec:discussion}

\subsection{Timeline of significant epochs for all historic observations}
\label{sec:timeline}

\begin{table*}
\centering
\caption{Chronicle of Significant Epochs for SN~2023ixf}
\setlength{\tabcolsep}{5pt}
\label{tab:timeline}                     
\begin{tabular}{l c c c c c c}          
\hline \hline
Measurement           &  Phase (or Phase Range ) &  JD &  Reference   \\

\noalign{\smallskip} \hline \noalign{\smallskip}
Epoch of Explosion                  &  --                  &   2460083.315         &   1           \\
Epoch of Discovery                  &  0.9\,d              &   2460084.227        &   7           \\
Epoch of Polarization Point A      &  1.4\,d              &                        &   4, This paper  \\
Epoch of Shock Breakout             &  2.2\,$\pm$0.2\,d    &                        &   This paper  \\
                                    &  2.75\,$\pm$0.3\,d   &                        &   3           \\
Epoch of X-ray Detection (NuSTAR)    &  3.9\,d       &                        &   6           \\
Epoch of X-ray Detection (Swift-XRT) &  4.25\,d &               &   3, 10           \\
Change in the profile of the Narrow $\rm H\alpha$ feature &   4.4\,d   &      &      2   \\    
Peak in UV Luminosity (\textit{blue peak})      &  4.5\,$\pm$0.5\,d    &      &   3, This paper  \\
Peak in He\,II line flux            &  4.5\,$\pm$0.5\,d    &                        &   3   \\
Transition in PA of Polarization    &  4.4 -- 4.6\,d (starts at 3.5\,d)   &                        &   4, This paper    \\
Epoch of Polarization Peak B      &  6.4\,d              &                        &   This paper  \\
Emergence of Broad $\rm H\alpha$ absorption  &  7\,d  &                        &   1, 2, 5, 8    \\
End of Flash-Ionisation Phase       &   8\,d               &                        &   1, 5, 8 \\
Peak of Soft X-ray Emission           &   10\,d             &                       &   3, 9, 10 \\
Disappearance of Intermediate-Width Lorentzian Features    &  $>$\,10\,d    &  &  1         \\
Peak in Luminosity of redder bands (\textit{red peak})   &  16\,d  &                        &   This paper  \\
Emergence of Broad HV $\rm H\alpha$ absorption  &  16\,d  &                        &   This paper  \\
Epoch of Millimeter Non-Detections (230 GHz)                         &   2.6 -- 18.6\,d      &  & 11 \\

Epoch of Radio Detection (10 GHz)                         &   29.2\,d      &  & 12 \\
\noalign{\smallskip} \hline \noalign{\smallskip} 
\end{tabular}
\newline
(1) \citet{2023teja}
(2) \citet{2023smith};
(3) \citet{2024zimmerman};
(4) \citet{2023vasylev};
(5) \citet{2023galan};
(6) \citet{2023grefenstette};
(7) \citet{2023ixfitagaki};
(8) \citet{2023bostroem};
(9) \citet{2023chandra};
(10) \citet{2023panjkov};
(11) \citet{2023berger};
(12) \citet{2023matthews}
\end{table*}

In its infancy, SN~2023ixf showed a rapid evolution spearheaded by the appearance of an increase in ionization in the early flash-ionization phase, accompanied by a rapid ascent in the early UV flux. Additionally, the increase in the color temperature to approximately 35,000 K until around 2.2 days (see Figure~\ref{fig:bollc}) signifies that the shock breakout was delayed and occurred within a confined, dense CSM. Our estimate for the epoch of shock breakout differs slightly from that of \citep{2024zimmerman}. This variation arises because they use the blackbody radius and the shock velocity approximated from the photospheric velocity to compute the extent of the dense CSM. The epoch of peak UV luminosity (and the bolometric peak) at 4.5\,d is synonymous with the peak of the \ion{He}{2} line flux of \citet{2024zimmerman}, which traces the strength/flux of the flash-ionization features. The change in the line profile of the narrow $\rm H\alpha$ feature and the drop in its strength seen in the high-resolution spectroscopy of SN~2023ixf \citep{2023smith} also happens around 4.4\,d, and is synonymous with UV peak. This indicates that the thermal heating and ionization continued beyond the shock breakout (2.2\,d). The emergence of a CDS \citep{2009chugai} within the post-shock CSM and the decelerated SN ejecta is probably contributing to the photo-ionization and prolonged heating observed in the flash-ionized features of SN~2023ixf. During the breakout phase, the radius of thermal emission remained relatively constant at (2.0\,$\pm$\,0.2)\,$\rm\times\, 10^{14}$\,cm (or 13\,$\pm$\,1 AU), indicating the location where thermal radiation originates from within the dense CSM \citep{2011chevalier}. This radius, derived from blackbody fits to UV-Optical-NIR data, is typically smaller (as it generally forms deeper) than the radius of photospheric emission ($\rm\tau\,\sim \,2/3$) \citep{2011moriya} and the surface of last scattering of the dense CSM. 

The first detection of SN~2023ixf in X-rays from \textit{NuSTAR} on $\sim$\,4\,d showed a large column density of absorption, consistent with arising from a shocked dense CSM \citep{2023grefenstette}. However, the next epoch of X-ray observations at 11\,d and 13\,d \citep{2023grefenstette, 2023chandra} exhibited a substantial decline in the column density of absorption. SN~2023ixf showed the emergence of broad P-Cygni of H$\rm \alpha$ at 7\,d, and the end of the flash-ionized phase lasted $\sim$\,8\,d. We see the appearance of the broad HV absorption of H$\rm \alpha$ at 16\,d in synonymity with the \textit{red peak} in our multi-band light curves in Section~\ref{sec:risetime}. The intermediate-width Lorentzian features from CSM interaction disappeared in the spectra around 16 -- 18\,d \citep{2023smith}. SN~2023ixf was not detected at millimeter wavelengths from 2.6 -- 18.6\,d \citep{2023berger}. SN~2023ixf was eventually detected in radio wavelengths rather feebly after 29.2\,d \citep{2023matthews}. The timeline of significant epochs for SN~2023ixf is tabulated in Table~\ref{tab:timeline} for reference.

\subsection{Progenitor of SN~2023ixf}
\label{sec:prog}

Numerous works on SN~2023ixf have estimated the progenitor mass, mass-loss rate, and extent of the CSM encompassing the progenitor. Pre-explosion imaging at the site of SN~2023ixf through \textit{HST} and \textit{Spitzer} revealed a point source consistent with an RSG enshrouded by a large amount of dust \citep{2023soraisam, 2023jencson, 2024neustadt}. However, no counterpart was discovered in UV or X-rays \citep{2023basu, 2023matsunaga, 2023kong, 2023panjkov}. There is, however, a disparity in the estimates of progenitor mass from the pre-explosion imaging revealing estimates in 2 broad ranges, i.e. (9 -- 14 $\rm M_{\odot}$, \citealp{2023kilpatrick, 2023pledger, 2023vandyk, 2024neustadt}, and (17 -- 22 $\rm M_{\odot}$, \citealp{2023jencson, 2023soraisam, 2023niu, 2023qin}. Our numerical hydrodynamical modeling best matched a ZAMS progenitor mass of 10 $\rm M_{\odot}$ for SN~2023ixf having a radius of 470 $\rm R_{\odot}$. Only other work that performed hydrodynamical modeling of the complete light curve until the early nebular phase indicated a 12 $\rm M_{\odot}$ progenitor \citep{2024bersten}. The short-plateau nature of SN~2023ixf indicates a relatively lower ejecta mass, which is also reflected in its steep plateau decline rate. Furthermore, the considerable blueshifted offset observed in H$\rm \alpha$ ($\sim$\,3,000 $\rm km\ s^{-1}$) during the early phase reinforces this deduction, indicating an escalated degree of stripping undergone by the 10 $\rm M_{\odot}$ progenitor of SN~2023ixf.
A summary of the various estimates of the progenitor mass and the extent and the mass-loss rate of the dense CSM is shown in Figure~\ref{fig:compparams}.

Pre-explosion observations of the progenitor of SN~2023ixf revealed variability in the mid-IR and near-IR observations from \textit{Spitzer} and ground-based telescopes \citep{2023kilpatrick, 2023soraisam, 2023jencson}. However, despite this variability, there is no indication of pre-SN outbursts in the pre-explosion imaging \citet{2023jencson, 2023hiramatsu, 2024ransome}, nor any signs of variability in the optical spectrum \citep{2023dong, 2024ransome, 2024neustadt}, which likely denies the existence of episodic mass loss in SN~2023ixf. This lack of pre-explosion outbursts, coupled with the presence of an asymmetric confined CSM (see section~\ref{sec:csmgeometry}), suggests that the progenitor of SN~2023ixf likely had a rapid rotation and/or underwent a pre-SN interaction with a binary companion \citep{2014smith, 2023matsuoka}. Such an interaction would have led to an enhanced mass loss in the lead-up to the explosion and drive a significant asymmetry in the observed CSM, as inferred from our polarization observations. SN~2023ixf is thus a low-mass RSG progenitor showcasing a multi-faceted CSM geometry arising from enhanced mass loss during its twilight years. The presence of the two-zone CSM in SN~2023ixf (see Section~\ref{fig:mesamodel}) supports the notion of an increased mass-loss rate of the RSG progenitor of SN~2023ixf as it evolved towards its explosion.

\subsection{Characteristics of circumstellar material around SN~2023ixf}
\label{sec:csmprog}

SN~2023ixf showed several signs of an interaction with CSM both photometrically and spectroscopically. The hydrodynamical modeling in Section~\ref{sec:modelmesa} emphasized the presence of a confined dense CSM responsible for the origin of the flash-ionization features, steep rise in UV flux, bolometric luminosity and temperature, and an extended low-density CSM, which led to the late-phase UV excess and the clumpy features around H$\rm \alpha$. This brings forward the argument that the progenitor of SN~2023ixf had an enhanced wind that developed shortly before the explosion, leading to the delayed shock breakout.

The modeling also revealed that the true extent of the confined CSM is roughly 5\,$\times$\,$\rm 10^{14}$ cm (33 AU), more significant than the nearly-flat radius of thermal emission during the early SN evolution in Section~\ref{sec:boltemp}. Our estimates align well with the estimates from comparison with CMFGEN models \citep{2023galan}, high-cadence early spectroscopy \citep{2023bostroem}, early light curve modeling \citep{2023hiramatsu} and pre-discovery photometry close to the explosion \citep{2024li}. Assuming a wind-velocity of 10 $\rm km s^{-1}$, the confined CSM is characterized by a wind-like structure with a mass-loss rate of $\sim$\,10$^{-2}$ $\rm M_{\odot}\ yr^{-1}$ which is consistent with estimates derived using comparison with CMFGEN models \citep{2023galan} and narrow emission features \citep{2024zimmerman}. The mass-loss rate indicates a progenitor star in the eruptive phase, which could lie anywhere between 10$^{-2}$ -- 0.1 $\rm M_{\odot}\ yr^{-1}$ \citep{smith_2017}. The absence of SN~2023ixf detections in mm-wavelengths by \citet{2023berger} within the first 4 days, suggests a mass-loss rate of $\gtrsim$\,10$^{-2}$ $\rm M_{\odot}\ yr^{-1}$, aligns with our modeling estimates and is likely influenced by a strong dominance of free-free absorption in the dense CSM along our line-of-sight.


The first detection of SN~2023ixf in X-rays from \textit{NuSTAR} on $\sim$\,4\,d displayed a large column density of absorption, consistent with arising from a shocked confined dense CSM \citep{2023grefenstette}. However, the next epoch of X-ray observations at 11\,d and 13\,d \citep{2023grefenstette, 2023chandra} exhibited a substantial decline in the column density of absorption arising from 3\,$\times$\,10$^{-4}$ $\rm M_{\odot}\ yr^{-1}$\citep{2023grefenstette}. The disparity between mass-loss rate estimates from X-ray observations and other measurements is likely due to the X-rays probing the interaction with the low-density extended CSM, whereas the others are probing the confined CSM. The lower mass-loss rates inferred from X-ray observations are consistent with our findings of interaction with a low-density extended CSM and our mass-loss rate estimate of $\sim$\,10$^{-4}$ $\rm M_{\odot}\ yr^{-1}$) from light curve modeling. The non-detection of SN~2023ixf in mm-wavelengths by \citet{2023berger} after the first 4 days hints towards a weak synchrotron emission from the low-density extended CSM \citep{2023hiramatsu}.

\subsection{Asphericity in the CSM / Outer Ejecta: Asymmetric confined dense CSM and Aspherical Shock front/extended CSM}
\label{sec:csmgeometry}

The temporal evolution of polarization from close to the explosion (1.4\,d) up until the middle of the plateau phase suggests significant departures from global sphericity for SN~2023ixf, which encompasses the confined CSM, outer ejecta and quite likely the extended CSM.

\subsubsection{Early polarization evolution: Point A}
\label{sec:peaka}

Our initial polarization observation (Point A; $p,\sim$\,1.1\%) at 1.4\,d post-explosion aligns with the appearance of flash-ionization lines in the spectral sequence of SN~2023ixf, a consequence of its interaction with the confined dense CSM. Given that the interaction with the dense CSM predominantly influences the luminosity during this phase, the inferred polarization during this period reflects the aspherical geometry of the dense CSM at a PA of $\sim$\,150\,$^{\circ}$. Early epoch polarization values are consistent with the findings of \citet{2023maund} and \citet{2023vasylev}. Since the shock breakout doesn't happen until 2.2\,d, the shock-powered SN ejecta lies within the compact dense CSM at 1.4\,d and indicates an asymmetric and confined dense CSM surrounding the progenitor of SN~2023ixf. During this phase, the narrow emission of $\rm H\alpha$, arising from the pre-shock CSM, shows a blueshifted and asymmetric feature \citep{2023smith}, potentially caused by the plane of the dense CSM aligning across our line of sight, obscuring the redshifted portion of the CSM.

Owing to the asphericity of the confined dense CSM and the arguments above, we consider the similarity of the CSM around SN~2023ixf to PTF11iqb \citep{2015smith} and SN~1998S \citep{2000gerardy}, which displayed an equatorial geometry of their CSM (i.e. a disk/torus). Similar to PTF11iqb and SN~1998S, SN~2023ixf exhibited a bright UV peak (see Section~\ref{sec:compsne}), flash-ionization features (see Section~\ref{sec:specflash}) and show Point A in polarization (PA\,$\sim$\,150$\rm^{\circ}$, see Section~\ref{sec:pol}) within the first four days. However, this phase in SN~2023ixf in similar in comparison to SN~1998S ($\sim$\,8 d) and longer than that of PTF11iqb ($\sim$\,3 d) \citep{2024galan}. This likely indicates the confined CSM in SN~2023ixf has a large radial extent, i.e., $\sim$\,30\,AU, Section~\ref{sec:modelmesa}). Assuming an oblate geometry for the CSM, the polarization at Point A corresponds to a major-to-minor axis ratio of $\sim$\,1.2 \citep{1991hoflich}.


\begin{figure*}
\centering
	 \resizebox{0.8\hsize}{!}{\includegraphics{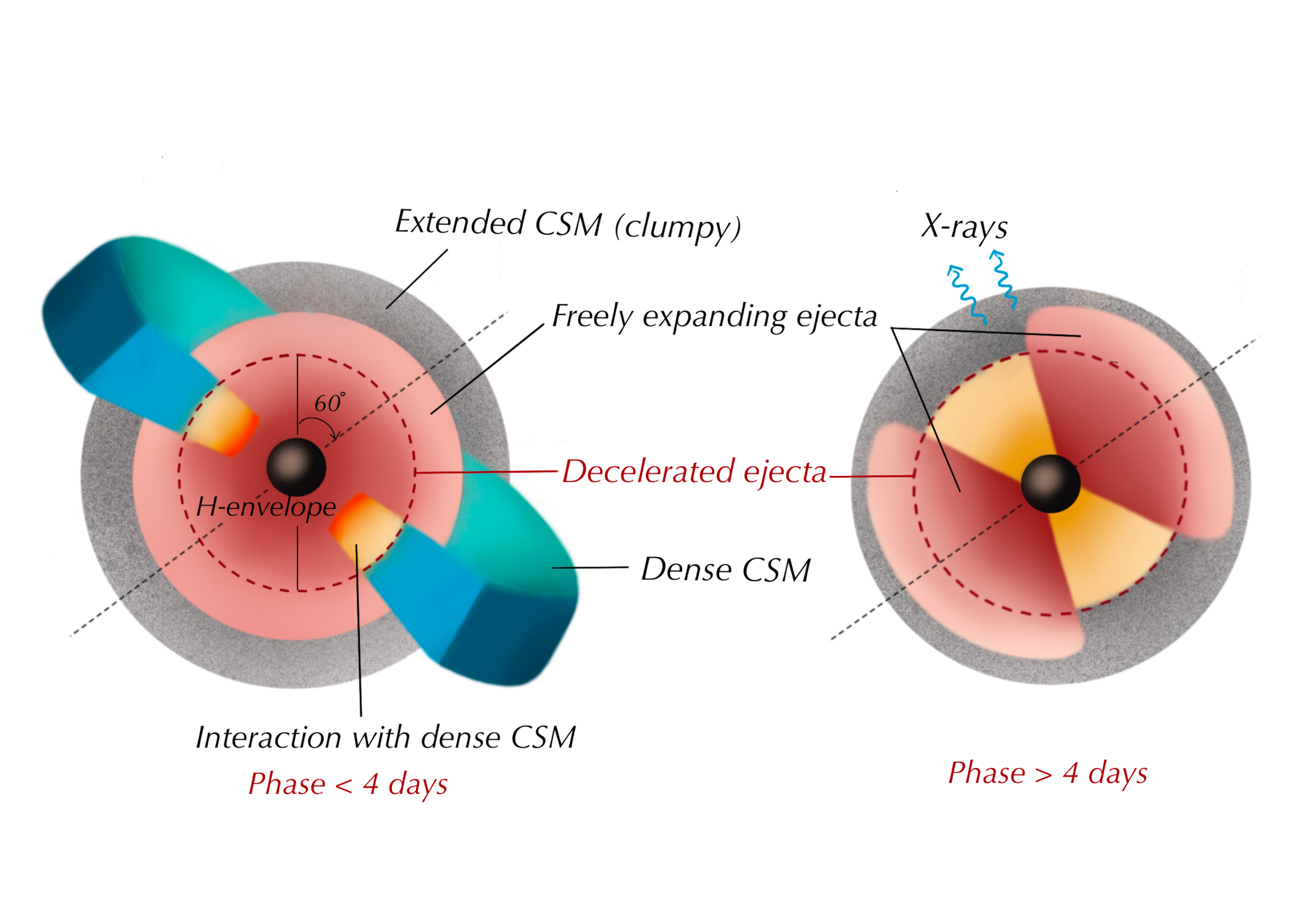}}
    \caption{Schematic illustration of the geometry (as viewed face-on by the observer) proposed for the CSM around the progenitor of SN~2023ixf depicted by its equatorial dense CSM and a clumpy, low-density extended CSM. The illustration on the left highlights the flash-ionized phase resulting from interaction with the dense CSM. On the right, we see an aspherical shock front arising after shock breaking out from the confined dense CSM and the continued interaction with the low-density extended CSM. CSM shows a projected major-to-minor axis ratio of 1.2, assuming an oblate geometry.}
    \label{fig:schematic}
\end{figure*}

\subsubsection{Photospheric phase polarization evolution: Peak B}
\label{sec:peakb}

We observe a shift in the imaging polarization evolution at 4.4\,d, suggesting a transition in the primary source of continuum luminosity, which was earlier dominated by interaction with the dense CSM. The aforementioned epoch aligns with the peak ionization and UV flux (refer to Table~\ref{tab:timeline} for the timeline); this PA change indicates an alternate source of polarized light. Moving forward, we observe a polarization rise to another point at 6.4\,d (Peak B; $p\,\sim$\,0.54\%). The polarization is likely a result of the recombination emission from an aspherical hydrogen envelope or additional luminosity from interaction with an asymmetric CSM. Following Peak B, we see a steady decline in polarization while the PA continued evolving slowly across the early plateau phase, falling to as low as $\sim$\,0.11\% on $\sim$\,22\,d. The epoch of polarization minima coincides with the changing slope during the plateau phase ($s1\ \rightarrow\ s2$). Our next epoch of imaging polarization at $\sim$\,28\,d sandwiched by two epochs of spectropolarimetric observations (at 27.3\,d and 30.9\,d), which showed an increase in polarization with the PA changing $\sim$\,50$\rm^{\circ}$ from the polarization minima at 22\,d. The early evolution of SN~2023ixf until the middle of the plateau phase showcases the dominant axis in the CSM-outer ejecta, which is illustrated in the q-u plane in Figure~\ref{fig:qupol}. Such a preferred alignment onto a dominant axis is a marker of global axisymmetry \citep{2017tanaka}. Since Point A and Peak B arise from a roughly axisymmetric geometry of the CSM and outer ejecta, the SN is viewed from an angle of inclination of 60 degrees to the plane of the sky. Larger viewing angles will require more significant asymmetries since the polarization will scale as $\rm \propto P(equatorial)\,sin (2\theta)$, where $\theta$ is the angle measured from the pole \citep{1991hoflich, 2001wang}. We explore the origin of Peak B from two potential scenarios outlined below: 

\begin{itemize}
    \item \textit{Interaction with an extended CSM which is asymmetric}: There is a clear indication that SN~2023ixf also harbors a low-density extended CSM due to visible signatures of intermediate-width features until 10\,d. In addition, we observe clumpiness in the extended CSM through the $\rm H\alpha$ line profile from 10 -- 32\,d, which could cause polarized emission. It is also reflected in the requirement for an extended CSM in the hydrodynamical modeling of the UV excess spanning up to the early nebular phase. However, the above effects are likely not strong enough to account for Peak B polarization unless the extended CSM exists as dense clumps in an effervescent zone as prescribed by \citet{2023soker}. In addition, \citep{2023smith} noted that the narrow emission from $\rm H\alpha$ migrates towards a narrower symmetric feature past $\sim$\,4.4\,d, indicating that the pre-shock CSM it is likely originating from, is symmetric considering its line geometry. We also observe the effects of UV excess from the interaction with the extended CSM interaction span until the nebular phase; however, the polarization drops drastically to a minimum at 22\,d. Additionally, the gradual evolution of the PA during the photospheric phase further weakens the argument that interaction with the extended CSM is the sole cause of this polarized light. 

    \item \textit{Asphericity in the hydrogen envelope}: We see the emergence of the broad $\rm H\alpha$ P-Cygni in SN~2023ixf at 7\,d, which coincides closely with the epoch of Peak B in polarization. This likely indicates that the polarization might arise due to the asphericity in the hydrogen envelope of the SN ejecta since the photosphere would reside in the hydrogen envelope during the photospheric phase. Additionally, we observe the appearance of a broad high-velocity absorption feature blueward of H$\rm \alpha$ and H$\rm \beta$, alongside an absorption feature associated with their P-Cygni profiles. The PA change of $\sim$\,90 degrees from Point A to Peak B emphasizes that the faster-moving ejecta (Peak B) lies perpendicular to the equatorial CSM (Point A). The switch in PA was also promptly observed between 3.5 and 4.6\,d in the spectropolarimetric observations of \citep{2023vasylev} wherein the PA of H$\rm \alpha$ at 3.5\,d becomes the PA of the continuum at 4.6\,d. This likely means that such a transition occurred earlier for Balmer line emission than for the continuum along the CSM-outer ejecta dominant axis. We hence firmly believe that the source of the dual broad Balmer absorption features and polarization Peak B arises from the aspherical shock front in SN~2023ixf, induced by its interaction with the equatorial dense CSM. We explain this further in the following subsection.
\end{itemize}

\subsubsection{Aspherical shock structure in SN~2023ixf}
\label{sec:schematic}

Consider the case of an RSG undergoing an explosion surrounded by a torus-shaped dense CSM across the equatorial plane along with a clumpy low-density extended CSM as shown in Figure~\ref{fig:schematic}. The shock traversing across the equatorial plane would be decelerated by the confined dense CSM, whereas the shock moving along the higher latitudes will expand freely owing to a low-density CSM, causing an aspherical shock front. Hence, such a geometry would likely allow for two broad absorption features in the spectra of SN~2023ixf arising from the decelerated ejecta and the freely expanding SN ejecta (see Figure~\ref{fig:halpha}). The decelerated ejecta broke out from the dense CSM along the equatorial latitudes at 2.2\,d. However, the unimpeded, freely expanding ejecta along the polar latitudes required it to traverse the lateral extent of the decelerated ejecta to become discernable later in the observations of SN~2023ixf. It is reasonable to anticipate a counterpart in the emission from the freely expanding ejecta. The distinct visibility of the HV absorption in our spectra implies a component of the freely expanding ejecta moving toward our line of sight. Consequently, this orientation would mask the redward emission, as it would be hindered by its motion away from our line of sight and potentially be obscured by the equatorial decelerated ejecta. When the broad HV absorption feature emerges, we observe a faint signature of extended red edge of the H$\rm \alpha$ emission feature around $\sim$\,16\,d (see Figure~\ref{fig:halpha}). 


An aspherical shock structure is crucial for prolonging the survival time of dust grains, resulting in time-varying extinction as was presented in the pre-discovery photometry of SN~2023ixf, $<$\,1 day from the explosion \citep{2024li}. In addition, the aspherical shock front is consistent with a prolonged and weak initial breakout flash in SN~2023ixf and a rising luminosity produced as ejecta flung out after breakout, expanding, and cooling. The asymmetric nature of the dense CSM also explains the low X-ray luminosity of SN~2023ixf in comparison to interacting SNe II \citep{2023panjkov}, in tandem with the highly asymmetric SN~2017hcc, which exhibited a bright optical-IR emission but a weak X-ray emission \citep{2022chandra}. \citet{2023panjkov} described that the stacked X-ray \textit{Swift-XRT} flux arises from a two-component bremsstrahlung emission consisting of a heavily absorbed hotter component and a less-obscured cooler component, which fits the description of an aspherical shock front in SN~2023ixf. 

\citet{2023grefenstette} and \citet{2023chandra} highlighted the discrepancy in shock breakout velocities derived from X-ray observations in contrast to the estimates based on optical observations by \citet{2023galan} and \citet{2024zimmerman}. X-rays in CCSNe are predominantly produced through shock interaction with nearby CSM, and the geometry of the CSM plays a significant role in determining the probability of X-rays escaping and their detectability in an SN. Specifically, an asymmetric CSM structure, like a disk or toroid, facilitates the escape of X-rays along higher latitudes, in contrast with a spherical CSM, where their escape would be impeded. In addition, the mass-loss rate derived by \citet{2023grefenstette} implies lower optical depth ($<$\,1) to electron-scattering in contrast to the presence of Lorentzian features in the optical spectral sequence until 10\,d. The asymmetry in the dense CSM and the aspherical shock structure interacting with the extended CSM likely drove the origin of the X-rays from a region distinct from the UV-Optical-NIR emission. They manifested as lower shock velocity measured by \citet{2023grefenstette, 2023chandra} using X-rays compared to the H$\rm \alpha$ velocity from our spectral sequence. 

The primary source of optical emission is driven by the shock decelerated by the equatorial CSM ($\sim$\,8,500 $\rm km s^{-1}$), forming the primary P-Cygni Balmer profile. In contrast, the X-ray observations by \textit{NuSTAR} and \textit{Chandra} at epoch 11 -- 13 d likely originate from the faster shock ($\sim$\,13,500 $\rm km s^{-1}$) arising from the freely expanding ejecta interacting with the low-density extended CSM, also leading to the HV Balmer absorption. 
A weak-synchrotron emission arising from interaction with the low-density extended CSM \citep{2023hiramatsu} is consistent with the lack of detections in mm-wavelengths by \citep{2023berger} post 4 days. The mass-loss rate estimates from X-ray observations are consistent with our estimate for the extended CSM (see Section~\ref{sec:modelmesa}. However, \citet{2023chandra} estimated that the radial evolution of X-ray emission past 13\,d evolved as $\rm t^{-1}$ is consistent with a wind profile of $\rm r^{-2}$ whereas our hydrodynamical modeling estimated a steeper wind profile of $\rm r^{-3}$ for the extended CSM, indicating a slight disagreement (see Section~\ref{sec:modelmesa}). However, it is important to acknowledge that neither the 1-D hydrodynamical modeling nor a single component X-ray emission fully captures the complexities of asymmetry in the CSM/ejecta of SN~2023ixf as observed through its polarization evolution. Initial attempts towards a two-component X-ray emission from \citep{2023panjkov} attribute it to the asphericity in SN~2023ixf, potentially explaining its underluminous nature. Future studies with 2-D/3-D hydrodynamical modeling with the wealth of information on SN~2023ixf will be necessary to address these details and further our understanding of its progenitor system.


\subsection{Geometry of the Inner Ejecta / He-Core: Peak C}
\label{sec:peakc}

Post the spectropolarimetric observations at 31\,d, we see an increased polarization in the next epoch of imaging polarimetric observation at 68\,d (0.44\,\%), which lies closer to the end of the plateau phase. There's a slight rise in polarization towards the 3rd peak (Peak C; $p\,\sim$\,0.5\%) at 79.2\,d, which lies during the transition from the plateau phase. As SN~2023ixf evolved into the nebular phase, the polarization showed a gradual decline in evolution. This decline is likely attributed to the expansion of the inner He-core, resulting in a reduction of its optical depth to electron scattering \citep{2006leonard}, consequently leading to a decrease in polarization. However, we also see an evolution of the PA past Peak C from $\sim$\,18\,$^{\circ}$ to $\sim$\,106\,$^{\circ}$, indicating that the scenario is more complex. The evolution of SN~2023ixf past the middle of the plateau phase until the early nebular phase aligns across alternate dominant axes showing axisymmetric inner ejecta / He-core (see Figure~\ref{fig:qupol}). 

We observe signs of asymmetry in the early nebular phase spectroscopy in the form of redshifted excess in H\,$\rm \alpha$ and [\ion{Ca}{2}] $\rm \lambda \lambda$7291,\,7324 at around +5,000 $\rm km\ s^{-1}$ in Figure~\ref{fig:nebularlines}. Additionally, a dual-peaked axisymmetric profile of the [\ion{O}{1}] doublet, blueshifted by 1500 $\rm km \ s^{-1}$, is also observed in SN~2023ixf during the early nebular phase in Figure~\ref{fig:nebularlines}. It is important to note that these signatures of asymmetry are distinct from the blueshifted excess seen close to the emission peak of the line profiles of H\,$\rm \alpha$ and [\ion{Ca}{2}] $\rm \lambda \lambda$7291,\,7324 which is likely a result of dust formation (see Figure~\ref{fig:nebularlines}). In the classification of [\ion{O}{1}] line profiles in CCSNe, as proposed by \citet{2009taubenberger}, the double-peaked structure of the [\ion{O}{1}] lines exhibiting axisymmetric blueshifted and redshifted peaks is associated with a toroidal distribution of oxygen. Such excitation of oxygen is expected from an asymmetric distribution of $^{56}$Ni \citep{2000gerardy} when observed from approximately 60 to 90 degrees with respect to the symmetry axis \citep{2008maeda, 2024fang}. The line profile of [\ion{O}{1}] doublet in the early nebular phase conforms to that picture, echoing the polarization signature detected during the early nebular phase. The effect of asymmetric distribution of the $^{56}$Ni has also been proposed for these deviations from asphericity \citep{2006Chugai, 2011Dessart}. In the case of SN~2004dj, \citep{2005chugai} argued that the bipolar ejection of $^{56}$Ni led to the asymmetric excitation of hydrogen within a spherical hydrogen envelope due to the shock propagation not being purely radial. In such a scenario, a rapidly but smoothly changing PA might be expected during the transition from the photospheric phase to the nebular phase, during which $^{56}$Ni is uncovered. This could explain the PA rotation observed in SN~2023ixf from 68\,d up until 140\,d. Such an ejection of $^{56}$Ni is hypothesized to stem directly from bipolar explosions in CCSNe resulting from mildly rotating RSG cores \citep{2005chugai}. 

All the facets concerning the geometry described above are compatible with a jet-driven explosion, where the jet breaks out from the inner core into the hydrogen envelope \citep{2001wang}. This would lead to an aspherical shock structure, not due to the interaction with the dense equatorial CSM but the asymmetric ejection of $^{56}$Ni. CCSNe are often associated with jets that puncture the outer layers or get choked by the hydrogen envelope \citep{2011couch}. There are a few SNe II, like SN~2012aw \citep{2014brajesh}, SN~2013ej \citep{2021nagao} and SN~2017gmr \citep{2019nagao}, which show an early onset of rise in polarization before the end of the plateau phase during the photospheric phase which is indicative of asymmetry in the hydrogen envelope, probably indicating a choked jet in the hydrogen envelope. The onset of polarization rise in SNe II appears to be linked to their luminosity, ejecta velocity, and $^{56}$Ni mass. CCSNe with higher luminosity, higher ejecta velocity, and increased $^{56}$Ni production exhibit more pronounced and prolonged aspherical characteristics \citep{2024Nagao}. This suggests that a higher likelihood of asphericity, possibly resembling a bipolar explosion leading to a jet-like formation, could be a crucial factor in the explosion mechanism of highly energetic SNe like SN~2023ixf. 

An intriguing observation is that SN~2023ixf shows two preferred dominant axes: a) along the CSM-outer ejecta, where the SN evolved during the early phase until the middle of the plateau phase, and b) along the inner ejecta/He-core, where the SN evolved during the late-plateau and early nebular phase. In other words, we see a change in PA of $\sim$45 degrees from Peak B to Peak C in SN~2023ixf, suggesting that the irregularities in the explosion structure are not aligned with the orientation of the confined CSM. Therefore, if a jet-driven explosion drove SN~2023ixf, the rotation of the progenitor did not drive the jet. A similar change in PA was noticed in the polarization of SN~1987A (see Figure~\ref{fig:pollc}) from the photospheric phase to the early nebular phase, indicating a potential avenue for further investigation if hydrogen-rich SNe could be powered by jets. The $R$-band imaging polarization evolution and spectropolarimetric observations of SN~2023ixf from close to the explosion (1.4\,d) up until the early nebular phase (140\,d) suggest significant departures from global sphericity in various facets of the evolution of SN~2023ixf, which encompasses the confined CSM, SN ejecta (possible the extended CSM), and the inner core.


\section{Summary}
\label{sec:summary}

SNe II are the most common type of CCSNe and yet harbor several mysteries regarding the late-stage evolution of its massive star progenitors, resulting in considerable observational heterogeneity. We present a comprehensive investigation of multi-wavelength photometry, optical spectroscopy, polarimetry, and spectropolarimetric observations of SN~2023ixf until the early nebular phase. We highlight the major results below:

\begin{itemize}
    \item {\bf Photometric evolution:} SN~2023ixf showed rise times of 4.5\,d (\textit{blue peak}) arising due to CSM interaction and 16\,d (\textit{red peak}) arising from the SN ejecta. SN~2023ixf shows a early plateau decline rate (s1) of $2.70^{+0.48}_{-0.49}$ $\rm mag\,(100\,d)^{-1}$ and a late-plateau decline rate (s2) of $1.85^{+0.13}_{-0.14}$ $\rm mag\,(100\,d)^{-1}$, resembling fast-declining SNe II. The plateau length of SN~2023ixf is 75\,d, towards the shorter end of SNe II. SN~2023ixf is one of the brightest SN IIP/L ever observed in UV with a peak UVW1 magnitude of $\sim$\,--20 mag.
    
    \item {\bf Slow photospheric evolution and Distinctive $\rm H\alpha$ profile:} We infer a delayed development of metal features in the spectral sequence of SN~2023ixf, hinting at the ejecta cooling slower than a normal SNe II, possibly due to CSM interaction. The weaker absorption in the P-Cygni profile of $\rm H\alpha$ suggests ongoing interaction during the plateau phase, reminiscent of Type IIL SNe. Post 16\,d, the $\rm H\alpha$ and $\rm H\beta$ are characterized by a high-velocity broad absorption feature at 13,500 $\rm km\ s^{-1}$ in addition to the clumpy P-Cygni profile with an absorption minimum at 8,500 $\rm km\ s^{-1}$. 
    
    \item {\bf Confined CSM}: The early discovery and classification allowed for the coverage of the SN just after 1\,d wherein we observe an initial rise in the blackbody temperature evolution, blueward rise of UV colors, and order-of-magnitude rise in UV flux at a nearly constant radius of evolution asserting the delayed shock breakout due to a confined dense CSM in SN~2023ixf.

    \item {\bf Hydrodynamical light curve modeling:} Using STELLA to perform hydrodynamical light curve modeling of SN~2023ixf, we estimated a progenitor mass of 10 $\rm M_{\odot}$ with a radius of 470 $\rm R_{\odot}$ and an explosion energy of $2\times 10^{51}~\mathrm{erg}$ and 0.06 $\rm M_{\odot}$ of $^{56}\,\mathrm{Ni}$. The inferred properties are not a unique solution since degeneracies exist in the modeling. The early UV excess was best modeled by a confined dense CSM spanning from the tip of the progenitor to $5 \times 10^{14}\,\mathrm{cm}$ arising from a mass-loss rate of 10$^{-2}$ $\rm M_{\odot}\ yr^{-1}$ and a $r^{-2}$ density structure. The late-plateau UV excess was modeled by an extended CSM spanning $5 \times 10^{14}\,\mathrm{cm}$ to $10^{16}\,\mathrm{cm}$ with a mass-loss rate of 10$^{-4}$ $\rm M_{\odot}\ yr^{-1}$ and a $r^{-3}$ density structure. A wind-velocity of 10 $\rm km\ s^{-1}$ was assumed for computing the mass-loss rates. The presence of the two-zone CSM in SN~2023ixf with 2 orders of magnitude difference in mass-loss rate suggests an increased mass-loss rate of the progenitor as it neared its explosion.

    \item {\bf Asphericity in CSM, ejecta, and the He-core:} The temporal evolution of ISP-corrected $R$-band polarization in SN~2023ixf showed three distinct peaks: Point A [$(1.09 \pm 0.05)\%$, $(153.4\pm 0.3)^{\circ}$] at 1.4\,d, Peak B [$(0.54 \pm 0.06)\%$ at 6.4\,d, $(60.3\pm 1.1)^{\circ}$] and Peak C [$(0.48 \pm 0.05)\%$, $(16.9\pm 0.8)^{\circ}$] at 79.2\,d. The observations indicate the presence of an asymmetric confined dense CSM, an aspherical shock front, and clumpiness in the low-density extended CSM, along with an aspherical inner He-core in SN~2023ixf. The inference on an aspherical shock front also stems from the dual broad absorption features seen in the blue wing of the Balmer features starting 16\,d. Overall, SN~2023ixf displayed two dominant axes, one along the CSM-outer ejecta and the other along the inner ejecta/He-core, showcasing the independent origin of asymmetry during the early and late evolution.

    \item {\bf Signs of Molecular CO / Dust formation:} The flattening in the $K_{s}$-band light curve and the attenuation of the red-edge of H$\rm \alpha$ post 125\,d indicates early onset of molecular CO and dust formation in SN~2023ixf similar to SN~2017eaw and SN~1987A. However, our dataset lacks an NIR spectrum during the nebular phase to confirm the presence of molecular CO.
    
\end{itemize}

SN~2023ixf, the third closest CCSNe observed in the 21st century, merits deeper exploration to comprehensively analyze its high-cadence observations across the electromagnetic spectrum. Conducting late-phase nebular spectroscopy holds promise in illuminating the intricate, asymmetric structure of the CSM surrounding SN~2023ixf and providing insights into the properties of its synthesized dust. We anticipate that the aspherical shock structure may manifest in the observations at radio wavelengths, revealing evidence of both freely expanding and decelerated shocks. The interaction of the faster shock with the low-density CSM is expected to generate a rapid rise in the radio light curve, while the interaction with the equatorial dense CSM may result in a slower increase over the subsequent months. These distinct radio signatures prompt further observational follow-up to validate the findings outlined in this article.


\section{Software and third party data repository citations} \label{sec:cite}

\vspace{5mm}
\facilities{HCT: 2-m, GIT: 0.7-m, KT: 1.5-m, Bok: 2.3-m, Seimei: 3.8-m, Iriki: 1-m, Oku: 0.51-m, ICSP: 0.61-m, Nayoro: 0.36-m, Swift (UVOT)}

\software{astropy \citep{astropy:2013, astropy:2018, astropy:2022}, emcee \citep{2013PASP..125..306F}, IRAF \citep{93_tody}, HEASoft \citep{2014ascl.soft08004N}, matplotlib \citep{Hunter:2007}, pandas \citep{mckinney-proc-scipy-2010, reback2020pandas}, numpy \citep{harris2020array}, scipy \citep{2020SciPy-NMeth}, Jupyter-notebook \citep{Kluyver2016jupyter}, seaborn \citep{Waskom2021}}. 

\section*{acknowledgments}
We thank the anonymous referee for their comments which helped in the readability of the manuscript. We thank the IAO, Hanle, CREST, and Hosakote staff, who made the observations from HCT possible. The facilities at IAO and CREST are operated by the Indian Institute of Astrophysics, Bangalore. The GROWTH India Telescope (GIT) is a 70-cm telescope with a 0.7-degree field of view, set up by the Indian Institute of Astrophysics (IIA) and the Indian Institute of Technology Bombay (IITB) with funding from Indo-US Science and Technology Forum and the Science and Engineering Research Board, Department of Science and Technology, Government of India. It is located at the Indian Astronomical Observatory (IAO, Hanle). We acknowledge funding by the IITB alumni batch of 1994, which partially supports the operation of the telescope. Telescope technical details are available at GROWTH-India Website\footnote{\url{https://sites.google.com/view/growthindia/}}. This research is based partly on data obtained through the KASTOR (Kanata And Seimei Transient Observation Regime) program. The Seimei telescope at the Okayama Observatory is jointly operated by Kyoto University and the National Astronomical Observatory of Japan (NAOJ), with assistance provided by the Optical and Infrared Synergetic Telescopes for Education and Research (OISTER) program funded by the MEXT of Japan. The authors thank the TriCCS developer team (supported by the JSPS KAKENHI grant Nos. JP18H05223, JP20H00174, and JP20H04736 and by NAOJ Joint Development Research). 

TJM is supported by the Grants-in-Aid for Scientific Research of the Japan Society for the Promotion of Science (JP24K00682, JP24H01824, JP21H04997, JP24H00002, JP24H00027, JP24K00668) and the Australian Research Council (ARC) through the ARC's Discovery Projects funding scheme (project DP240101786). KM acknowledges support from the JSPS KAKENHI grant JP20H00174 and JP24H01810. DKS acknowledges the support provided by DST-JSPS under grant number DST/INT/JSPS/P 363/2022. GCA thanks the Indian National Science Academy for support under the INSA Senior Scientist Programme. BK acknowledges support from the ``Science \& Technology Champion Project'' (202005AB160002) and from two ``Team Projects" -- the ``Top Team'' (202305AT350002) and the ``Innovation Team'' (202105AE160021), all funded by the ``Yunnan Revitalization Talent Support Program".

Numerical computations were in part performed by the PC cluster at the Center for Computational Astrophysics (CfCA), National Astronomical Observatory of Japan. This research made use of \textsc{RedPipe}\footnote{\url{https://github.com/sPaMFouR/RedPipe}} \citep{2021redpipe}, an assemblage of data reduction and analysis scripts written by AS. This research also made use of the NASA/IPAC Extragalactic Database (NED\footnote{\url{https://ned.ipac.caltech.edu}}), which is funded by the National Aeronautics and Space Administration and operated by the California Institute of Technology. 

The work used Swift/UVOT data reduced by P. J. Brown and released in the Swift Optical/Ultraviolet Supernova Archive (SOUSA). SOUSA is supported by NASA's Astrophysics Data Analysis Program through grant NNX13AF35G. This work has also used software and/or web tools obtained from NASA's High Energy Astrophysics Science Archive Research Center (HEASARC), a service of the Goddard Space Flight Center and the Smithsonian Astrophysical Observatory. This work was also partially supported by a Leverhulme Trust Research Project Grant. 


\bibliography{_SN2023ixf}{}
\bibliographystyle{aasjournal}
\appendix

\begin{figure}[hbt!]
	 \resizebox{\hsize}{!}{\includegraphics{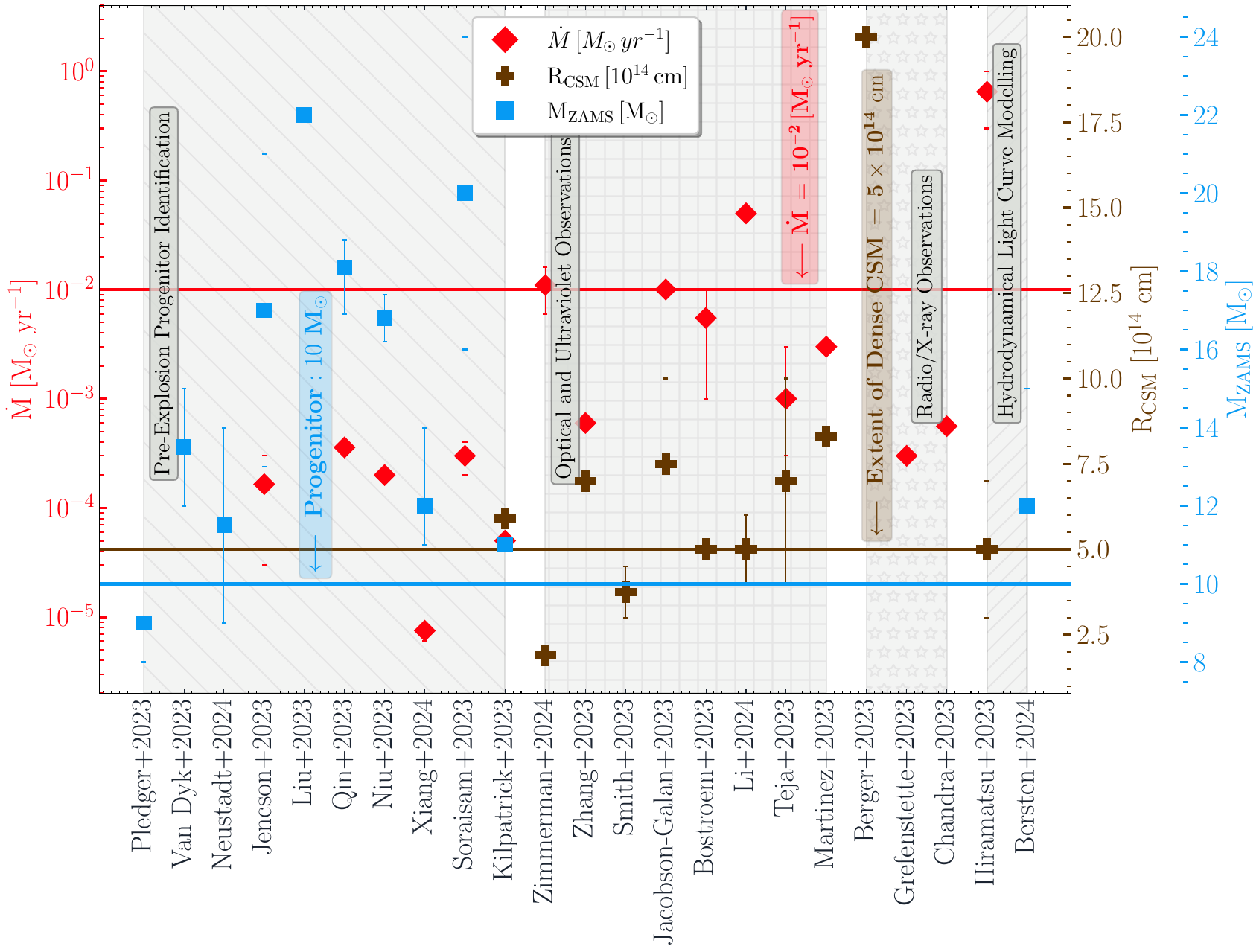}}
    \caption{Compilation of estimates for the parameters associated with the progenitor of SN~2023ixf, including progenitor mass, CSM extent, and mass-loss rate. The techniques used to obtain the plotted estimates are indicated with distinct hatch patterns, such as pre-explosion progenitor identification, optical and UV observations, radio/X-ray observations, and hydrodynamical light curve modeling. The horizontal solid lines, shown in different colors, represent our estimates derived from hydrodynamical light curve modeling. Various wind velocities were assumed in different studies when estimating mass-loss rates, which are noted here for clarity: 10 $\rm km\ s^{-1}$ \citep{2023teja, 2023jencson, 2023bersten, 2024neustadt, 2024zimmerman}, 50 $\rm km\ s^{-1}$ \citep{2023galan, 2023soraisam, 2023qin, 2023kilpatrick}, 70 $\rm km\ s^{-1}$ \citep{2024xiang}, 115 $\rm km\ s^{-1}$ \citep{2023smith, 2023niu, 2023berger, 2023hiramatsu, 2023martinez}, and 150 $\rm km\ s^{-1}$ \citep{2023bostroem}.}
    
    \label{fig:compparams}
\end{figure}

\end{document}